\documentclass[lettersize,journal]{IEEEtran}
\usepackage[ruled,linesnumbered]{algorithm2e}    
\usepackage{booktabs}                            
\usepackage{centernot}                           
\usepackage{colortbl}                            
\usepackage{graphicx}                            
\usepackage{hhline}                              
\usepackage{lipsum}                              
\usepackage{mathtools}                           
\usepackage[all]{nowidow}                        
\usepackage{optidef}                             
\usepackage{pgfplotstable}                       
\usepackage{subcaption}                          
\usepackage[binary-units]{siunitx}                             
\usepackage{substr}                              
\usepackage{tikz}                                
\usepackage{tcolorbox}                           
\usepackage{xcolor}                              
\usepackage{circledsteps}

\definecolor{light-gray}{gray}{0.968}

\usepackage{xspace}                              
\usepackage{glossaries}
\usepackage{multirow}
\usepackage{float}
\usepackage{url}
\usepackage{pgf}
\usepackage{wrapfig}

\usepackage{pifont}

\usepackage{epsfig}
\usepackage{endnotes}
\usepackage{textcomp}
\usepackage{comment}
\usepackage{enumitem}
\usepackage{microtype}
\usepackage{tabularx}

\newif{\ifrevision}
\revisionfalse

\newcommand{\hypertargetblue}[2]{\textbf{\textcolor{blue}{\hypertarget{#1}{#2}}}}

\newcommand{\re}[2][]{[R#2.#1]}

\def\BibTeX{{\rm B\kern-.05em{\sc i\kern-.025em b}\kern-.08em
    T\kern-.1667em\lower.7ex\hbox{E}\kern-.125emX}}

\newcommand{\blue}[1]{#1}

\newcommand{\BFS}{\texttt{BFS}}

\newcommand{\nbf}[1]{\vspace{5pt}\noindent\textbf{#1}}

\usepackage[breaklinks]{hyperref} 
\DeclareSIUnit{\nothing}{\relax}

\PassOptionsToPackage{breaklinks}{hyperref}

\begin{document}

\title{Securing Cloud File Systems with Trusted Execution}
\author{Quinn Burke,~\IEEEmembership{Graduate Student Member,~IEEE}, Yohan Beugin, Blaine Hoak, Rachel King, Eric Pauley,\\Ryan Sheatsley, Mingli Yu,~\IEEEmembership{Graduate Student Member,~IEEE}, Ting He,~\IEEEmembership{Senior Member,~IEEE}, Thomas La Porta,~\IEEEmembership{Fellow,~IEEE}, Patrick McDaniel,~\IEEEmembership{Fellow,~IEEE}
\thanks{\scriptsize
The views and conclusions contained in this document are those of the authors and should not be interpreted as representing the official policies, either expressed or implied, of the Combat Capabilities Development Command Army Research Laboratory or the U.S. Government. The U.S. Government is authorized to reproduce and distribute reprints for Government purposes not withstanding any copyright notation here on. This work was also supported in part by the National Science Foundation under award CNS-1946022. This work was also supported in part by the Semiconductor Research Corporation (SRC) and DARPA. \textit{(Corresponding author: Quinn Burke.)}}
\thanks{\scriptsize Quinn Burke, Yohan Beugin, Blaine Hoak, Rachel King, Eric Pauley, Ryan Sheatsley, and Patrick McDaniel are with the Computer Sciences Department, University of Wisconsin-Madison, Madison, WI 53706 USA (e-mail: qkb@cs.wisc.edu; ybeugin@cs.wisc.edu; bhoak@cs.wisc.edu; rachelking@cs.wisc.edu; epauley@cs.wisc.edu; sheatsley@wisc.edu; mcdaniel@cs.wisc.edu).}
\thanks{\scriptsize Mingli Yu, Ting He, and Thomas La Porta are with the Department of Computer Science and Engineering, The Pennsylvania State University, University Park, PA 16802 USA (e-mail: mxy309@psu.edu; tinghe@psu.edu; tfl12@psu.edu).}
}

\markboth{Journal of \LaTeX\ Class Files,~Vol.~14, No.~8, August~2021}%
{Shell \MakeLowercase{\textit{et al.}}: A Sample Article Using IEEEtran.cls for IEEE Journals}


\maketitle

\begin{abstract}
Cloud file systems offer organizations a scalable and reliable file storage solution. However, cloud file systems have become prime targets for adversaries, and traditional designs are not equipped to protect organizations against the myriad of attacks that may be initiated by a malicious cloud provider, co-tenant, or end-client. Recently proposed designs leveraging cryptographic techniques and trusted execution environments (TEEs) still force organizations to make undesirable trade-offs, consequently leading to either security, functional, or performance limitations. In this paper, we introduce \BFS{}, a cloud file system that leverages the security capabilities provided by TEEs to bootstrap new security protocols \blue{that deliver strong security guarantees, high-performance, and a transparent POSIX-like interface to clients. \BFS{} delivers stronger security guarantees and up to a $2.5\times$ speedup over a state-of-the-art secure file system. Moreover, compared to the industry standard NFS, \BFS{} achieves up to $2.2\times$ speedups across micro-benchmarks and incurs $<1\times$ overhead for most macro-benchmark workloads. \BFS{} demonstrates a holistic cloud file system design that does not sacrifice an organizations' security yet can embrace all of the functional and performance advantages of outsourcing.}
\end{abstract}


\section{Introduction}
\label{sec:intro}

Cloud file systems are a backbone of modern cloud infrastructure. Often used as the storage interface for personal cloud drives and enterprise server applications, they provide convenient and reliable access to shared file data. While advantageous for several reasons, storing file data in the cloud raises significant security and privacy concerns~\cite{takabi2010security}. 

Breaches of private user data and metadata, intellectual property theft, and ransomware campaigns have been shown to be particularly effective in cloud environments~\cite{continella_theres_2018,alspach_microsoft_2022,scaife_cryptolock_2016}, highlighting the need for better ways of protecting data stored in the cloud. Further, adversaries in cloud environments include not only co-tenants and end-clients, but even a malicious cloud provider. More sophisticated defenses are required to mitigate attacks initiated by a malicious cloud provider (e.g., host system call tampering)~\cite{chen2005non,van2019tale,cui2021emilia,khandaker2020coin,ports2008towards}; these are commonly denoted as \textit{host-interface attacks}.
Concretely, a trusted cloud file system must therefore provide: (1) confidentiality and integrity protection for all file data and metadata, (2) resilience against a variety of host-interface attacks, (3) support for canonical features like file sharing and policy management, and (4) practical performance.

Designing a cloud file system that simultaneously meets all of these requirements is a challenging task. Widely-used cloud file systems like Amazon's EFS or Google's Filestore~\cite{haynes_network_2015,ghemawat_google_2003,aws-efs,google-filestore} can deliver high-performance, but necessarily force organizations to simply trust that neither the cloud provider, nor any other privileged or unprivileged adversary, can or will maliciously access or modify file data or metadata stored on the remote hosts. And while recent efforts have leveraged cryptographic techniques and trusted execution environments (TEEs) to secure data, they still force organizations to make undesirable trade-offs and fail to deliver either sufficient security controls, feature support, or performance guarantees~\cite{baumann_shielding_2015,arnautov_scone_nodate,tsai_graphene-sgx_nodate,shinde_panoply_2017,priebe_sgx-lkl_2020,blaze_cryptographic_1993,djoko_nexus_2019}. This has
consequently prevented these designs from seeing wide adoption as a primary storage interface. Thus, the community lacks a suitable file system that strikes a good balance between real-world security, functional, and performance requirements.

In this paper, we introduce \BFS{}, a cloud file system that meets real-world security, functional, and performance requirements. \BFS{} leverages the security capabilities provided by TEEs~\cite{sabt_trusted_2015,mckeen_innovative_2013,kaplan_amd_2016,ngabonziza_trustzone_2016} to bootstrap new security protocols that grant four key properties: (1) confidentiality and integrity protection for all file data and metadata; (2) comprehensive protection against host-interface attacks; (3) secure and high-performance file sharing; and (4) extensible feature support. \BFS{} demonstrates that organizations need not sacrifice file system security to embrace the functional and performance advantages of outsourcing.

Accomplishing this requires addressing a range of challenges associated with request processing and data persistence. First, protecting confidentiality and integrity requires designing novel end-to-end protocols that can mitigate various known attacks with minimal overhead. We address this through \textit{data \& metadata isolation}, wherein we securely partition file system tasks across trusted and untrusted components to efficiently protect against tampering with data and metadata while in-flight, in-processing, and at-rest. Second, protecting against host-interface attacks requires a careful reconsideration of the host-interface design to be able to reason about and mitigate them. We address this through \textit{host-interface shielding}, wherein we design a simple, deterministic host-interface and develop mechanisms to protect against tampering with host-interface parameters or return codes. Lastly, providing secure and high-performance file sharing and extensible feature support requires a reliable but versatile cryptographic key management system that minimizes the risk of key compromise and has minimal performance overhead. We address this by \textit{offloading cryptographic work} to the TEE, where the TEE serves as a trusted key escrow that manages persistent encryption keys and negotiates ephemeral keys with clients as needed.

Our evaluation of \BFS{} examines the design trade-offs in meeting real-world security, functional, and performance requirements. We first perform a security analysis of \BFS{} against a broad set of adversaries within the network, in-memory, and on-disk. We then provide an implementation of \BFS{}, running in a live, cloud-like environment, and evaluate the performance across a series of Filebench-based~\cite{Tarasov2016FilebenchAF} micro- and macro-benchmarks. Our analysis juxtaposes \BFS{} against the industry standard NFS~\cite{haynes_network_2015,aws-efs,google-filestore} \blue{and a state-of-the-art SGX-based file system NeXUS~\cite{djoko_nexus_2019}. We demonstrate the \BFS{} delivers equivalent stronger guarantees than NeXUS and up to a $2.5\times$ speedup. We also show that, compared to NFS (with Kerberos encryption enabled), \BFS{} delivers up to $2.2\times$ speedups across micro-benchmarks and incurs $<1\times$ overhead for most macro-benchmark workloads. \BFS{} takes a holistic approach to cloud file system design, demonstrating that it is possible to deliver strong security, high performance, and client transparency.}

We contribute the following:
\begin{enumerate}
    \item An end-to-end design and implementation of \BFS{}; \BFS{} provides comprehensive confidentiality and integrity protection for all file data and metadata, shields the host-interface, enables secure and high-performance file sharing, and enables extensible feature support.
    \item A security analysis demonstrating the resilience of \BFS{} against a wide range of both known and new attacks in the network, in-memory, and on-disk. 
    \item A performance analysis demonstrating that \BFS{} can ensure stronger security guarantees while providing practical performance w.r.t. state-of-the-art systems.
\end{enumerate}

\section{Background}
\label{sec:background}

\subsection{Cloud File Systems}
Cloud file systems extend the file storage capabilities of local file systems (e.g., \textit{ext4}~\cite{cao_ext4_2006}) to a cluster of outsourced \textit{server} and \textit{storage} hosts (or \textit{nodes}) connected to \textit{clients} by a network~\footnote{This architecture falls under the umbrella of distributed file systems.}. Here, the file system is similarly composed of both global and per-file data structures that track the file system \textit{data} (e.g., file contents) and \textit{metadata} (e.g., file attributes and data locations). Server and storage nodes cooperate in organizing, storing, and retrieving data and metadata for clients under a shared file system; the storage nodes may be local (directly-attached) or remote (connected via a storage-area network or other network transport~\cite{breuer_enhanced_2000,legtchenko2017understanding}). To clients, the distributed nature of the file system is transparent; once mounted, the files presented under the mount point have the same access semantics as files stored on any local file system. Widely supported implementations of these principles include the Network File System (NFS)~\cite{haynes_network_2015}, Amazon's Elastic File System (EFS)~\cite{aws-efs}, and Google's Filestore~\cite{google-filestore}.

Conventional architectures typically follow a centralized client-server model~\cite{haynes_network_2015,shvachko_hadoop_2010,ghemawat_google_2003}. Here, clients issue file I/O requests on behalf of end-users or applications (whether executing on-premises or outsourced themselves) to a centralized server across a network; the server itself exposes a POSIX-like file interface for clients to access files under a shared namespace. In executing file operations, the server organizes the file data and metadata as fixed-sized \textit{blocks} across the storage nodes; the storage nodes expose a simple interface for the server to store and retrieve blocks (typically \SI{4}{K\byte} in size). Clients typically coordinate these tasks with server and storage nodes through \textit{remote-procedure call} (RPC) request and response messages.

\subsection{Trusted Execution Environments}
Trusted execution environments (TEEs) are hardware-based security primitives that isolate execution of mutually distrusting software components running on a shared host. The software components may be other tenants' user-level applications, a hypervisor, or other system software. TEEs also provide attestation capabilities, allowing remote clients to ensure the legitimacy of the code running on an endpoint with whom they are communicating.

TEEs accomplish this through \textit{access-mediation}, hardware-based complete mediation over designated protected (inside the TEE) and unprotected (outside the TEE) regions of physical memory, or additional \textit{CPU modes}, processor modes that restrict the scope of operations that particular software components may perform within their execution context~\cite{mckeen_innovative_2013}. As a result, code and data residing in the TEE is granted strong confidentiality and integrity protection even in the presence of malicious software or hardware external to the TEE. Mature TEE implementations offering these capabilities include \textit{Intel~SGX}~\cite{mckeen_innovative_2013}, \textit{AMD~SEV}~\cite{kaplan_amd_2016}, and \textit{ARM~TrustZone}~\cite{ngabonziza_trustzone_2016}.

\section{Security Model}
\label{sec:sec-model}

\begin{figure}[!t]
    \centering
    \includegraphics[width=0.475\textwidth]{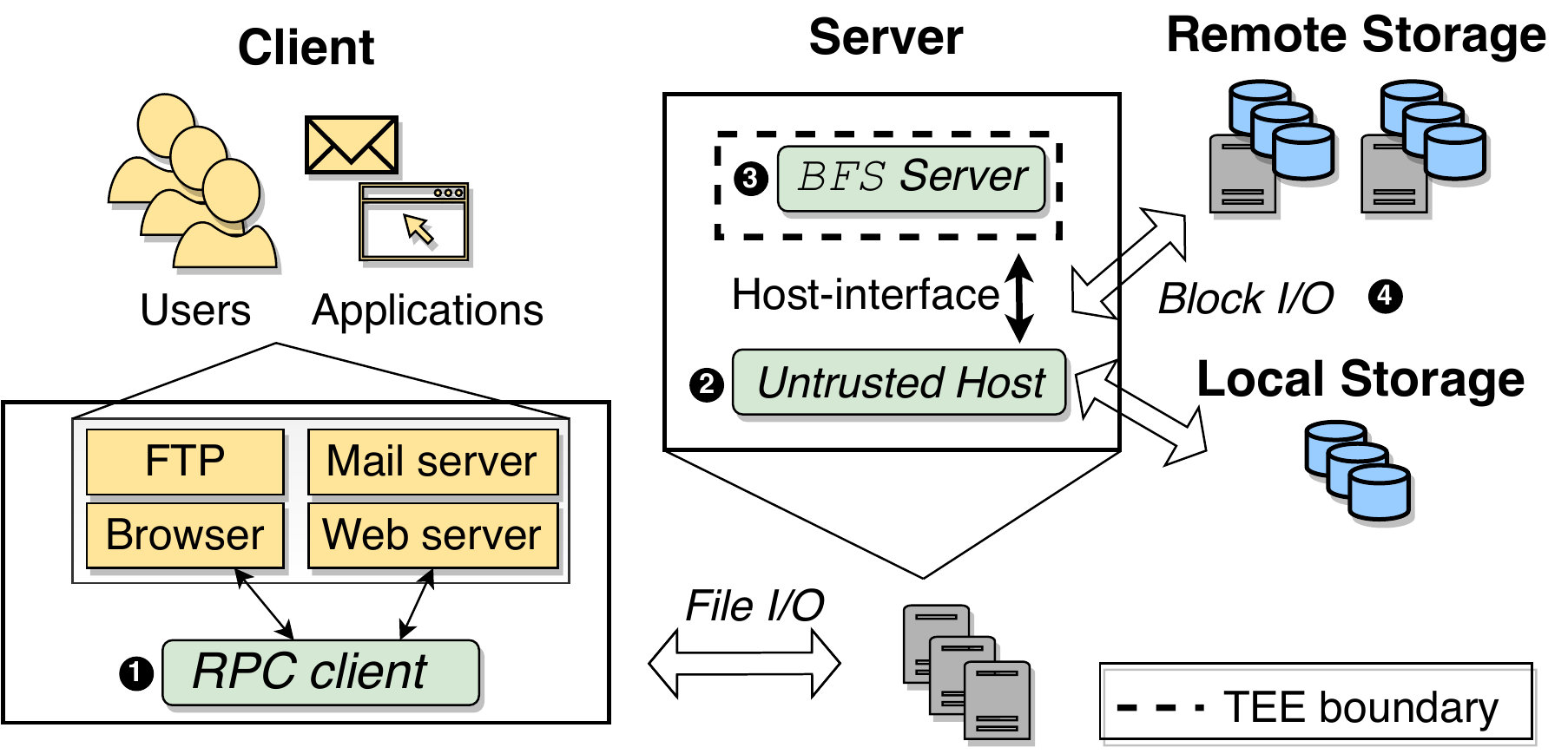}
    \caption{System components and workflow. \ding{202}\,Clients perform file I/O by having the \ding{203}\,untrusted host proxy RPC messages to the \ding{204}\,\BFS{} server and \ding{205}\,storage nodes.}
    \label{fig:bfs-components}
\end{figure}

\nbf{System Components.} We assume a centralized client-server model (see \autoref{fig:bfs-components})~\cite{depardon_analysis_nodate,haynes_network_2015}. The file system is orchestrated by five components: \textit{client}, \textit{network}, \textit{server}, \textit{TEE}, and \textit{storage}. On the frontend, the client software provides a file interface to either end-users (e.g., employees in an enterprise network) or applications (e.g., a company's web servers). The client communicates over the network to a TEE running on an outsourced server. The tasks at the server are handled by code running either inside of the TEE or outside---denoted hereafter as the ``\BFS{} server'' and ``untrusted host'', respectively. The storage backend consists of the outsourced local or remote storage nodes that receive commands to store or retrieve data in fixed-sized blocks. In executing file I/O requests, messages between the clients, \BFS{} server, and storage nodes are proxied by the untrusted host.

\nbf{Trust Model.} We consider an unmanaged deployment model, where the organization deploys and administers the file system. However, our design principles also extend to fully-managed deployments, where the cloud provider offers file storage as-a-service. We envision \BFS{} as a replacement for widely-used systems in either case~\cite{haynes_network_2015,aws-efs,google-filestore}. We therefore consider a client and TEE trusted components, and the network, untrusted host, and storage nodes untrusted. We assume the client trusts the TEE implementation.

\nbf{Threat Model.}\label{sec:threat-model} 
Our threat model is rooted in three key observations: both file data \textit{and} metadata have become high-yield targets for adversaries~\cite{scaife_cryptolock_2016,greschbach_devil_2012,cash2015leakage,maffei_privacy_2015,chen_titanium_2022}, host-interface attacks are a significant threat to TEE-based software~\cite{checkoway_iago_nodate,khandaker2020coin,van2019tale}, and weak or complex cryptographic key management increases the risks associated with key compromise~\cite{fumy1993principles,ateniese2006improved}.
As such, the system is subject to attempts to maliciously \textit{access}, \textit{corrupt}, \textit{swap}, \textit{replay}, \textit{reorder}, or \textit{drop} data sent between the clients, untrusted host, \BFS{} server, and storage nodes~\cite{kallahalla_plutus_2003,tsai_graphene-sgx_nodate,priebe_sgx-lkl_2020,arnautov_scone_nodate}. The untrusted host may abuse the host-interface---for example, by crafting malicious arguments or return values to hijack control-flow between client/storage and the TEE. And lastly, adversaries may attempt to steal the keys used to encrypt data on-disk.

In line with prior work, we consider denial-of-service, physical, and side-channel attacks out of scope (e.g., network-traffic analysis~\cite{fu2003active} and other TEE-based side-channel attacks~\cite{tsai_graphene-sgx_nodate}). We further discuss the limitations of extant TEEs in \autoref{sec:discussion:limitations}. Our threat model resembles those of recent TEE-based file systems, but differs in the wider range of attacks that we aim to address together---notably, swapping attacks, host-interface attacks, and key compromise.

\nbf{Security Requirements.} To meet real-world security requirements, the file system must therefore provide end-to-end confidentiality and integrity protection for both file data and metadata, protection against host-interface attacks, and a reliable cryptographic key management system that minimizes the risks associated with key compromise.

\section{Design Challenges}
\label{sec:design-challenges}

At surface-level, designing a file system that meets our security requirements may appear a trivial task: encrypt data, sanitize inputs, etc. However, designing an end-to-end solution is a much more nuanced endeavor. For example, while encrypting data suffices to protect confidentiality, there are security and performance trade-offs in deciding who has access to encryption keys and where encryption occurs.
We characterize the key challenges under three themes.

\label{dc:1}\nbf{C1.~Protecting confidentiality and integrity.}
Ensuring confidentiality requires new mechanisms that isolate all file data and metadata from untrusted components while in-flight, in-processing, and at-rest. Ensuring integrity requires being able to attest the authenticity and correctness of code and data while processing client requests. The central challenge here lies in deciding how to securely partition tasks across trusted and untrusted components. In particular, at the server, the trust and privilege levels of components need to be considered at a far more granular level than in conventional designs~\cite{haynes_network_2015,tsai_graphene-sgx_nodate}. Current TEE-based file systems still leave open several avenues for attack (e.g., expose metadata), and a simple port of a file server like NFS to a TEE runtime still leaves many security issues unresolved (e.g., key management). We must therefore develop a new set of end-to-end protocols that enable us to more sensibly reason about and mitigate attacks, with minimal overhead.

\label{dc:2}\nbf{C2.~Protecting the host-interface.} Mitigating host-interface attacks is a central challenge for cloud software~\cite{khandaker2020coin,van2019tale}. In our context, a malicious host or storage node may craft malicious arguments or return values to divert control-flow or cause other confidentiality and integrity violations. For example, valid, encrypted block data may be unknowingly swapped in place of that actually requested, before being delivered to the TEE from storage. Data encryption alone cannot defend against such attacks.
Further, reasoning about and mitigating them across large and complex host-interfaces has been shown to be infeasible~\cite{van2019tale,khandaker2020coin}; prior efforts provide support only for a limited set of defenses~\cite{shinde_besfs_nodate}. The typical TEE-based library operating system (libOS) model~\cite{tsai_graphene-sgx_nodate,shinde_panoply_2017} is therefore ill-fit for use here. To comprehensively protect against them therefore requires judicious host-interface design and techniques that consider how inputs from the host may affect higher-level file system semantics.

\begin{figure*}[!ht]
    \centering
    \includegraphics[width=\textwidth]{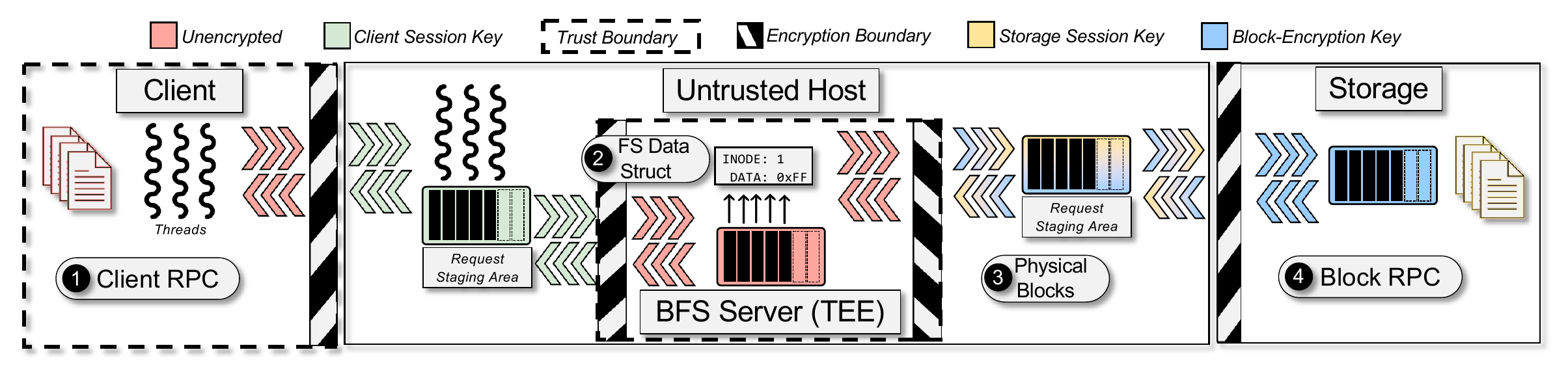}
    \caption{\BFS{} design. Clients request \ding{202}\,file-level I/O by communicating with the \BFS{} server through an RPC interface. The \BFS{} server handles the client requests by \ding{203}\,updating the file system data structures within the TEE. The \BFS{} server coordinates with the untrusted host to store and retrieve the underlying \ding{204}\,blocks on storage nodes through a \ding{205}\,block RPC interface.}
    \label{fig:bfs-overview}
\end{figure*}

\label{dc:3}\nbf{C3.~Supporting diverse file system features securely and efficiently.}
Cloud file systems are expected to support typical features like file sharing and high-level policy management~\cite{ateniese2006improved,wang2013robustness,vahldiek2015guardat,castro1999practical,alagappan_protocol-aware_nodate,shah2007auditing}. While various cryptographic techniques have been proposed to realize this, such approaches have significant practical limitations. For example, the typical, client-centric encrypt-then-upload model requires clients to support ad hoc cryptographic protocols and manage additional secrets. This increases the risks associated with key compromise (from lack of expertise, social engineering, or other human oversight). It complicates the semantics of file sharing; supporting a common application service like collaborative document editing is infeasible here. Moreover, it introduces performance limitations and additional constraints on feature support (e.g., supporting compliance auditing for an enterprise). Reconciling these concerns therefore requires a key management system that is reliable, versatile, and low-overhead.

\section{\BFS{} Design}
TEEs provide a unique opportunity to challenge the basic premise of prior cloud file system designs. However, while TEEs provide primitives to isolate and mediate access to sensitive data in memory, extending those guarantees beyond system memory to remote clients and persistent storage media is non-trivial. The challenges stem from the fact that TEEs are sandboxed environments and rely on the untrusted host (or other kind of supervisor) to proxy access to external resources like network cards. Some information must therefore be exposed to the host such that it correctly executes requests on behalf of the TEE. How to enable this capability securely and efficiently remains an open question.

Our central goal is therefore to seek out new abstractions that provide a more practical set of trade-offs. Our design is guided by three design principles:

\begin{itemize}

\item\label{dp:1}\textbf{Isolate data and metadata.} 
We use the strong security guarantees of TEE hardware to bootstrap new security protocols that protect the confidentiality and integrity of all file data and metadata while in-flight, in-processing, and at-rest.

\item \label{dp:2}\textbf{Provide shielding support.} 
We pivot on our isolation protocols to develop a comprehensive set of mitigations against host-interface attacks.

\item \label{dp:3}\textbf{Offload cryptographic work.}
We introduce an escrow-based key management system that leverages TEE capabilities to reduce the risks associated with key compromise and streamline feature support.
\end{itemize}

\subsection{Isolating Data and Metadata}
Conceptually, isolating data and metadata requires two tasks: deciding where file operations should execute and what the host-interface should look like. This is challenging for several reasons. First, both data and metadata are sensitive information, as they directly (through file contents, permissions, etc.) disclose private information about users and who they communicate with. They must therefore exist in plaintext only within the TEE (or client memory). Code running inside the TEE must then be able to understand the notions of directories and files to some extent, and code running outside should not be able to learn what the sensitive data is.

Second, guaranteeing the integrity of file I/O requests requires that the core file system logic (file operation handlers) be attestable by clients. Using a libOS or other POSIX wrapper library that deserializes client requests but then redirects them onto a local file system managed by the untrusted host precludes clients from being able to have assurance over how the file operation is actually implemented underneath.

Third, the decision of how to partition tasks as above directly impacts the granularity of the resulting host-interface. Opting for a libOS or wrapper library may reduce development efforts in porting core file system code to run within the TEE~\cite{tsai_graphene-sgx_nodate,shinde_panoply_2017,arnautov_scone_nodate}), but comes at the expense of an enlarged host-interface that then needs protection. Current defense efforts for libOSes provide support only for a limited set of attacks~\cite{shinde_besfs_nodate}. Such approaches also observe significant performance overheads, often $>10\times$ (and sometimes $>100\times$) end-to-end for local and remote clients~\cite{tsai_graphene-sgx_nodate,orenbach_eleos_2017,arnautov_scone_nodate,ahmad_obliviate_2018}.

Toward this, we introduce three abstractions: a \textit{trusted file system core}, \textit{secure I/O channels}, and a \textit{partitioned block layer}.

\subsubsection{Trusted File System Core}
In \BFS{}, the file operation handlers execute entirely within the TEE. As shown in~\autoref{fig:bfs-overview}, the \BFS{} server first consumes a buffered file or block RPC message from a queue located in unprotected memory, decrypts and deserializes it, then dispatches it to the appropriate file operation handler. Any outbound file or block RPC messages are then serialized, encrypted, and submitted through a similar queue in unprotected memory. Note that the file system has a metadata layout akin to \textsc{Unix}-based local file systems~\cite{cao_ext4_2006}, 
with a superblock, inode table, etc. Any data or metadata resident outside of the TEE is opaque to the untrusted host. And our design therefore reduces the host-interface size to only four functions: sending and receiving file and block RPC messages.

\subsubsection{Secure I/O Channels}
Bridging the clients on the frontend to the storage nodes on the backend then requires a secure transport layer. While standardized protocols like TLS provide means to realize this, the question here is what data can or should reside at the transport layer and above it.

\begin{figure}[!t]
    \centering
    \includegraphics[width=0.45\textwidth]{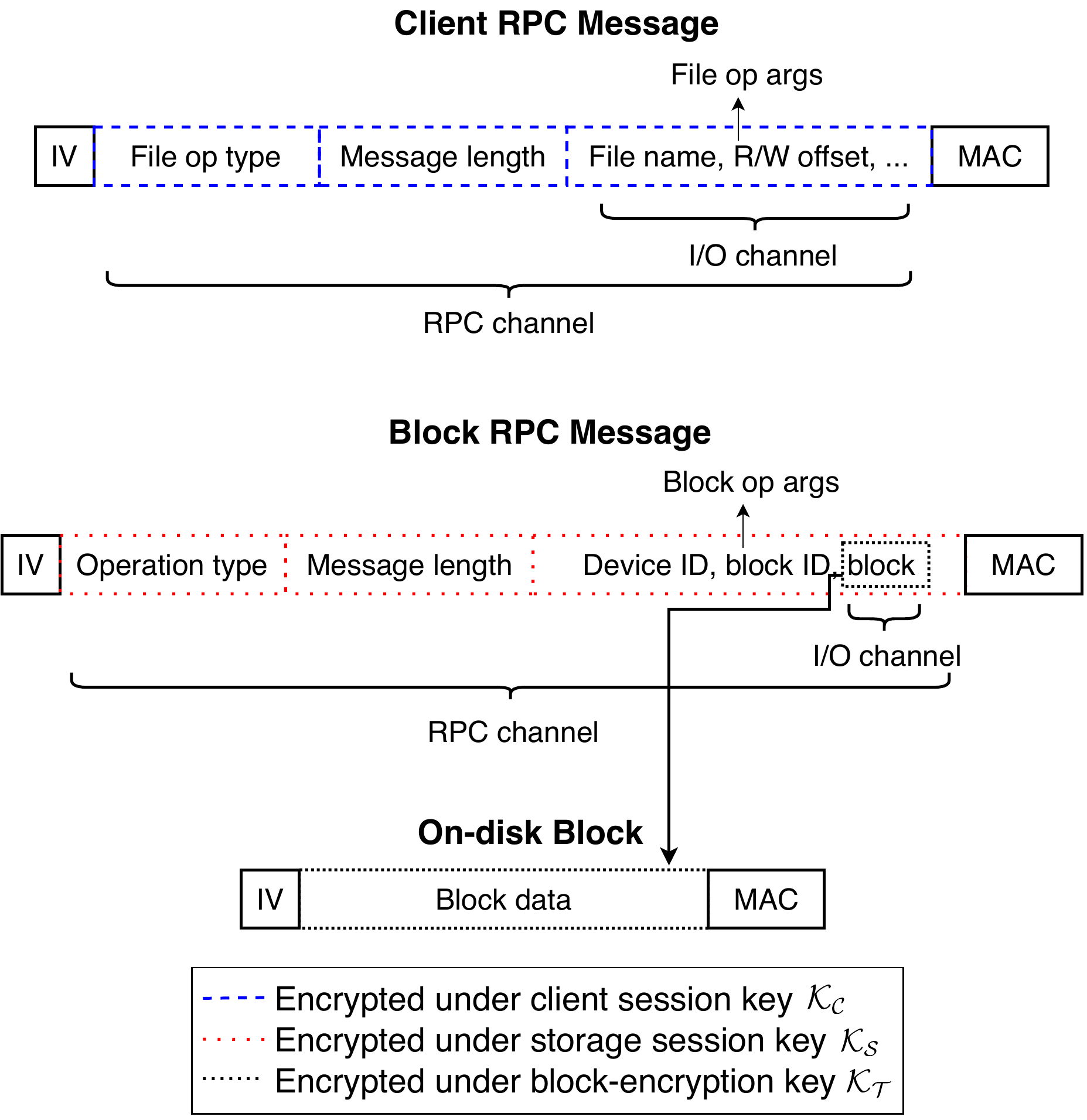}
    \caption{This message format captures the intuition behind our secure protocols for isolating file data and metadata.}
    \label{fig:bfs-message-format}
\end{figure}

We first distinguish between two distinct types of communication channels: I/O channels and RPC channels. As shown in~\autoref{fig:bfs-message-format}, I/O channels form logical connections between two endpoints. In contrast, RPC channels serve as the transport for I/O channels. I/O channels thus may contain sensitive data (file names, contents, R/W offsets, etc.) that must be kept secret from untrusted components, and we therefore require them to be terminated in the TEE. While RPC channels contain non-sensitive data (assuming an encrypted payload) that need not be kept secret, terminating the RPC channels in the untrusted host (which has been the de facto best practice) introduces vulnerabilities to host-interface attacks. We similarly require RPC channels to be terminated in the TEE; we defer further discussion on this to~\autoref{sec:shielding}.

File I/O requests from clients are therefore protected by encrypting and authenticating all I/O parameters under ephemeral key $\mathcal{K_C}$ before issuing them to the \BFS{} server. $\mathcal{K_C}$ is known only to them. MACs are computed over the request buffer and sequence numbers tracked by the client and \BFS{} server.

Block addresses (device ID/block ID pairs) must be exposed to the untrusted host and storage nodes such that they can correctly route and execute block I/O requests. We therefore treat plaintext block data as sensitive, but block addresses as non-sensitive. As detailed below, we encrypt block data prior to being marshalled into I/O requests. However, here block addresses may equally be stored in plaintext inside or outside the TEE. As an additional layer of integrity protection against network adversaries, block I/O requests are similarly encrypted and authenticated by the \BFS{} server and storage nodes under ephemeral key $\mathcal{K_S}$.

\subsubsection{Partitioned Block Layer}
The block layer is the exit point in the TEE where data must be prepared to be stored persistently on disk.
Blocks are first encrypted and authenticated by the \BFS{} server under a persistent block-encryption key $\mathcal{K_T}$; blocks are similarly decrypted in the TEE when retrieved from storage. The key is known only to the \BFS{} server, and therefore the block I/O channel is terminated only at the \BFS{} server.
After encryption, blocks are marshalled into (and unmarshalled from) block I/O requests by the \BFS{} server and delivered to storage nodes by the untrusted host.
We note that blocks may therefore be doubly-encrypted and authenticated: first as blocks (under $\mathcal{K_T}$), then as block RPC payloads (under $\mathcal{K_S}$). As an additional layer of protection, the block address is similarly authenticated by the \BFS{} server.

\subsubsection{Balancing Security and Performance} LibOSes have been central to TEE-based software development, but they are not a one-size-fits-all tool.

\nbf{Strong Isolation.} In \BFS{}, clients are presented a canonical POSIX file interface. We take a microkernel approach to the server design, providing a file-system-as-a-service that is attestable to clients and ensures the confidentiality, integrity, and freshness of client data and all code handling the data.

\nbf{Cutting Costs.} Yet, the significance of this design extends beyond simply that we protect metadata and prescribe a smaller host-interface. It enables us to more efficiently design integrity protection mechanisms. We implement blanket integrity protection for all files at the block layer rather than the file system layer (i.e., provide full-disk encryption capabilities without the downsides of current FDE methods). This enables us to avoid having to use ad hoc solutions for ensuring integrity---e.g., per-file hashes, which can be difficult to translate to block-level representations suitable for storage on disk~\cite{djoko_nexus_2019}.
It also offers performance advantages. It eliminates extraneous abstractions on the critical path to storage---like syscall interfaces, VFS layers, etc. Further, it avoids having to recompute costly checksums/hashes over large files for trivial changes (e.g., single-block updates)~\cite{zhang_end--end_nodate}.

\subsection{Providing Shielding Support}
\label{sec:shielding}
Prescribing a smaller host-interface is critical to being able to more easily mitigate host-interface attacks; we contrast this with libOS approaches that expose tens or hundreds of host-interface methods. In \BFS{}, 
state transitions at the host-interface are deterministic and predictable for all four message types. Unlike prior works, we can therefore exhaustively reason about how a malicious host may tamper with the interface parameters and return codes.
We introduce three additional abstractions: \textit{authenticated dispatch}, \textit{shielded block layer}, and \textit{guarded control transfer}.

\subsubsection{Authenticated Dispatch}
The entry point for client requests at the server is the RPC layer. While RPC systems have been well-studied, how to properly terminate an RPC channel in a TEE is an open question. Terminating RPC channels in the untrusted host (by simply hooking the functions running in the TEE to appropriate RPC handler stubs) has been key to accelerating I/O in TEE-based systems~\cite{orenbach_eleos_2017,tsai_graphene-sgx_nodate,google-asylo}. However, this approach directly exposes RPC opcodes to the untrusted host, and are therefore vulnerable to the untrusted host simply changing the opcodes to invoke arbitrary RPC handlers. For read-only interfaces, this can cause incorrect data to be returned to users or applications, and for read-write interfaces, this can cause mutations to the file system state to be incorrect.

The root of the problem stems from RPC interfaces containing handler functions with similar or identical function signatures. Consider a host-interface with methods for opening files and changing file permissions. An \texttt{open} operation has the signature \texttt{int open(const char *pathname, int flags)}, and a \texttt{chmod} operation has the signature \texttt{int chmod(const char *pathname, mode\_t mode)}, with \texttt{mode\_t} defined as the same integer type. With identical signatures, a malicious host can recast an \texttt{open} operation into a \texttt{chmod} operation, and the TEE will interpet the same (valid) I/O parameters in a different context. This would allow the host to induce a permissions change on a file.

\begin{figure}[!t]
    \centering
    \includegraphics[width=0.5\textwidth]{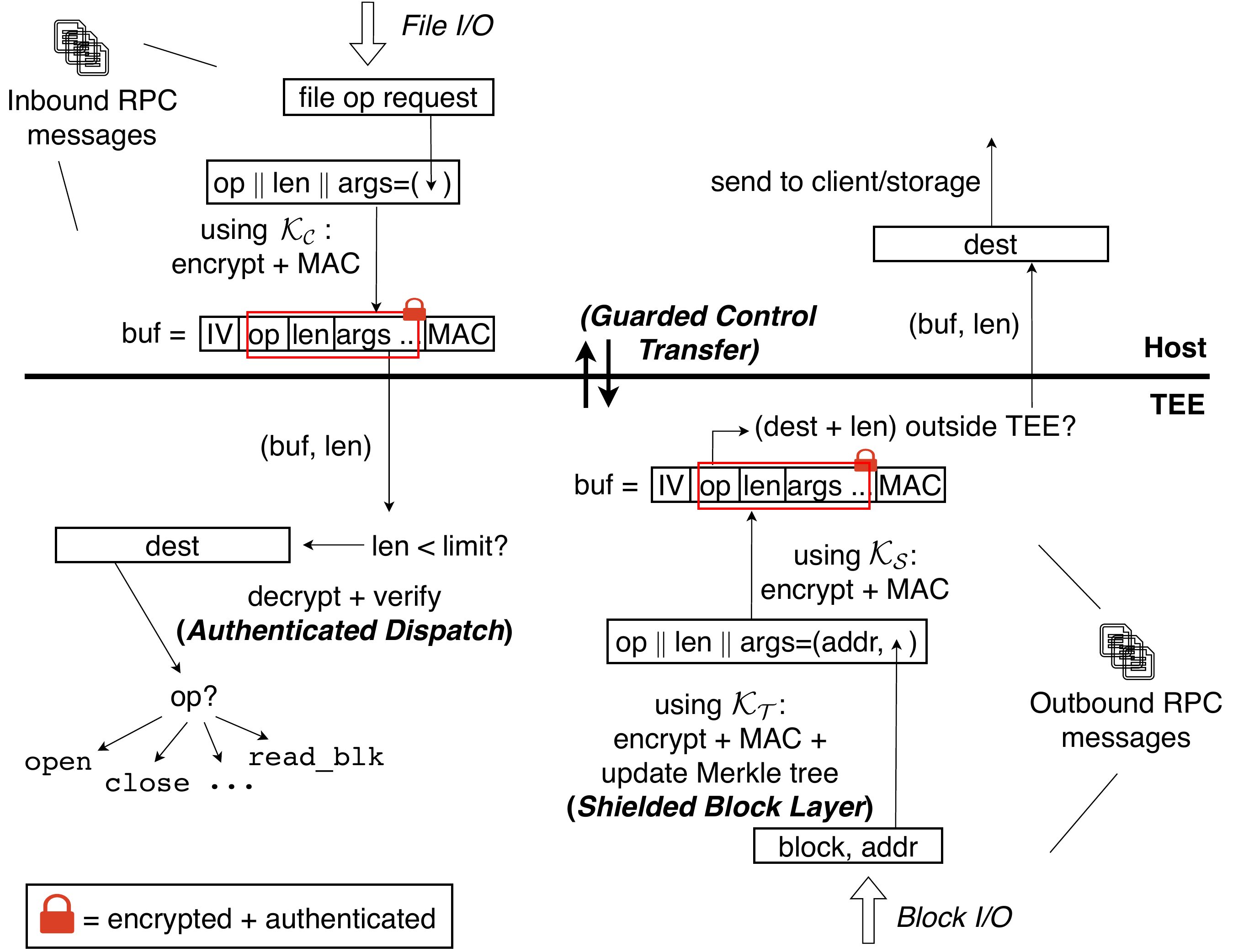}
    \caption{By authenticating RPC opcodes, using a tailored MAC construction and Merkle tree, and guarding control transfer, \BFS{} shields against host-interface attacks.}
    \label{fig:shielded-host-interface}
\end{figure}

In \BFS{}, we therefore consider RPC opcodes sensitive and terminate file RPC channels (in addition to I/O channels) in the TEE. All file RPC parameters are authenticated and verified by the \BFS{} server and clients before any file operation proceeds.
As shown in~\autoref{fig:shielded-host-interface}, once clients attest the TEE, this ensures controlled dispatch of file I/O requests: only the file operation requested by the client is invoked by the \BFS{} server. 
Note that we could alternatively delegate RPC tasks to the untrusted host by exposing, but still authenticating, the type code. However, we aim to limit the number of entry points into the TEE---keeping the interface size small, constant (w.r.t. the number of supported file operations), and deterministic. We also aim to reduce costs associated with crossing protection boundaries to perform integrity checks.

\subsubsection{Shielded Block Layer}
The exit point for requests on the backend is the block layer. In contrast to client RPC channels, block RPC parameters are considered non-sensitive because we require the untrusted host and storage nodes to handle persistence of blocks. We do not enforce similar restrictions on block RPC messages. However, the TEE stills need to be able to detect and respond to a malicious host that supplies corrupt, replayed, or swapped blocks. Towards this, we extend the calculation for block MACs to use the \textit{block address} as additional authenticated data (AAD) along with the block data. We then use a Merkle hash tree~\cite{merkle2001certified}, with the block MACs as the leaves, to prevent block replays/rollbacks and ensure the freshness of block data when retrieved by the TEE. The tree is stored on a separate region of the disk and read into protected memory on boot.

Like other file systems, our Merkle tree protects block correctness and freshness. However, the Merkle tree alone is still prone to second-preimage attacks and cannot prevent valid, encrypted blocks from being swapped in place of those actually requested before being delivered to the TEE from storage. In contrast to prior designs, our MAC construction therefore additionally prevents block swapping attacks.

\subsubsection{Guarded Control Transfer}
Our authenticated dispatch and shielded block layer mechanisms mitigate confidentiality and integrity violations resulting from maliciously crafted host-interface parameters. It remains to consider how to mitigate attacks resulting from malicious return codes supplied to the TEE by the host. While prior work has studied similar host-interface attacks to some extent~\cite{khandaker2020coin,van2019tale}, current mitigations address only a few specific attacks~\cite{shinde_besfs_nodate}.

The root of the problem lies in how return codes are typically handled in file systems. Standard practice in Linux for system calls like \texttt{read()} is to propagate return codes (both successes and errors) up the call stack from device drivers, through the block layer and file system layer, and back to clients~\cite{gunawi2008eio}. Note that the different software layers all use standard Unix \texttt{errno} codes. However, simply permitting the untrusted host to propagate arbitrary return codes would enable it to exploit vulnerabilities or weaknesses in the error-handling or decision-making logic in the file system or application code. We therefore need a mechanism to more rationally handle return codes.

The \BFS{} server intercepts return codes from the untrusted host and handles them in one of two ways. A return code of zero indicates a success, and the server proceeds. Any logical failures will be detected by the trusted file system core and either handled locally (e.g., retry block-write) or reported back to the client. Any other return code indicates a failure and is transformed into a generic I/O failure (EIO) before being reported back up the call chain to the client. False positive return codes will therefore be detected on a subsequent read/write via the Merkle tree. False negative return codes may only cause retries at the server (up to some limit) or generic I/O errors at the client; indeed physical I/O errors are typically handled transparently by the cloud provider~\cite{aws-ebs-sla,aws-failover}. In the absence of formally-verified file system or application code, this provides a hardened file system that raises the bar for attackers looking for control-flow exploits.

\subsection{Offloading Cryptographic Work}
\label{sec:key-mgmt}
Supporting a diverse set of features securely and efficiently has been a central challenge in secure file system design. Several decades of research have broadly focused on client-side encryption techniques (i.e., encrypt-then-upload) as a means for protecting outsourced file data~\cite{blaze_cryptographic_1993,kallahalla_plutus_2003,djoko_nexus_2019}. \blue{\ifrevision\hypertargetblue{r11}{\re[1]{1}}\fi/\ifrevision\hypertargetblue{r23}{\re[2]{3}}~\fi While shown to be useful in some contexts, such designs are ill-fit for typical usage patterns of cloud storage. Most notably, this requires complex, interactive, client-to-client protocols to perform simple tasks like sharing a file. This diverges from the server-focused, POSIX-based NFS model that most cloud applications are accustomed to.}
We take a different approach in \BFS{} by offloading as much cryptographic work to the \BFS{} server as possible. We introduce two abstractions: a \textit{trusted key escrow} and a \textit{shared persistent key}.

\subsubsection{Trusted Key Escrow \& Shared Persistent Key} In \BFS{}, we recognize the \BFS{} server as an extension of each client running on the outsourced server and appoint it as a trusted key escrow for clients. We now revisit the use of the block-encryption key ($\mathcal{K_T}$) and ephemeral client ($\mathcal{K_C}$) and storage ($\mathcal{K_S}$) keys; our security protocols are shown in~\autoref{tab:sec_proto}. We first distinguish between the notions of encryption for \textit{persistence} and for \textit{transport}: data is encrypted for persistence for the purpose of being stored on disk and encrypted for transport for the purpose of being sent in RPC messages.

The \BFS{} server encrypts blocks for persistence under key $\mathcal{K_T}$, shared by all clients. We note that $\mathcal{K_T}$ may represent a single key or a master key from which other persistent keys are derived (but shared by all clients). $\mathcal{K_T}$ is known only to the \BFS{} server and generated when the server first boots. Blocks are encrypted for transport in RPC messages to clients under a per-client session key $\mathcal{K_C}$. The key is negotiated when a client mounts the file system. Note that block RPC parameters are treated as non-sensitive (as blocks are internally shielded), thus we do not require a similar construction for $\mathcal{K_S}$ (it may or may not be ephemeral/shared among storage nodes).

\subsubsection{Key Maintenance} Using a trusted key escrow introduces additional challenges for bootstrapping the file system. For generating and storing persistent keys (like $\mathcal{K_T}$), prior work has relied on unique sealing keys burned into the processor hardware on the server. Yet, part of the advantage in outsourcing lies in the flexibility in service placement: the \BFS{} server may be migrated to a different machine due to third-party control-plane decisions, server failure, etc. Besides flexible placement, persistent keys must also be rotated occasionally to prevent attacks enabled by cryptanalysis; coupling the persistent key to the physical machine complicates this. When outsourcing, we therefore require more flexibility in how $\mathcal{K_T}$ is generated and stored.

In \BFS{}, $\mathcal{K_T}$ is machine-independent (i.e., initialized when the file system is formatted). We then use the unique sealing key of the TEE as a key-encrypting key to persist $\mathcal{K_T}$ on the current machine where \BFS{} is running. This provides hardware-backed persistence of the block-encryption key, without requiring any additional key service (third-party or otherwise), and while retaining data availability as the \BFS{} server is relocated to different physical machines. Moreover, it allows administrators to rotate persistent keys as often as necessary (without requiring a separate physical machine) and enables a seamless key transition period (by permitting incremental re-encryption of data under the new key). 

\subsubsection{Balancing Security, Performance, and Utility} 
The central challenge with supporting diverse requirements lies in how to efficiently manage encryption keys.
We highlight the advantages of our approach below.

\begin{figure}[!t]
\centering
\footnotesize
\fcolorbox{black}{white}{
\parbox{0.95\linewidth}{

\textbf{File I/O Messaging}
\begin{enumerate}
    \item $C \rightarrow T : \{m_1\}_{\mathcal{K_C}},MAC_{\mathcal{K_C}} (\{m_1\}_{\mathcal{K_C}},s_C)$ \hfill \textit{(client RPC request)}
    \item $T \rightarrow C : \{m_2\}_{\mathcal{K_C}},MAC_{\mathcal{K_C}} (\{m_2\}_{\mathcal{K_C}},s_{T,C})$ \hfill \textit{(client RPC response)}
\end{enumerate}

\textbf{Block I/O Messaging}
\begin{enumerate}
    \setcounter{enumi}{2}

    \item $T \rightarrow S : \{b_1\}_{\mathcal{K_S}},MAC_{\mathcal{K_S}} (\{b_1\}_{\mathcal{K_S}},s_{T,S})$ \hfill \textit{(block RPC request)}
    \begin{enumerate}
    \item $b_1 \leftarrow addr$ \hfill \textit{(read)}
    \item $b_1 \leftarrow addr,\{b\}_{\mathcal{K_T}},MAC_{\mathcal{K_T}} (\{b\}_{\mathcal{K_T}},addr)$ \hfill \textit{(write)}
    \end{enumerate}
    
    \item $S \rightarrow T : \{b_2\}_{\mathcal{K_S}},MAC_{\mathcal{K_S}} (\{b_2\}_{\mathcal{K_S}},s_S)$ \hfill \textit{(block RPC response)}
    
    \begin{enumerate}
    \item $b_2 \leftarrow addr,\{b\}_{\mathcal{K_T}},MAC_{\mathcal{K_T}} (\{b\}_{\mathcal{K_T}},addr)$ \hfill \textit{(read)}
    \item $b_2 \leftarrow ACK$ \hfill \textit{(write)}
    \end{enumerate}
    
\end{enumerate}
}
}
\caption{\BFS{} security protocols. $C$, $T$, and $S$ represent the client, \BFS{} server, and storage node. Sequence numbers are denoted by $s$.}
\label{tab:sec_proto}
\end{figure}

\nbf{Assessing Risks.} Entrusting the \BFS{} server with the persistent key enables us to overcome many of the security risks associated with conventional client-side encryption approaches. Indeed, (non-TEE-based) delegated key management has become a pivotal aspect of cloud services~\cite{aws-kms,chandramouli2013cryptographic}; TEEs provide a unique opportunity to capitalize on both the security and performance advantages of delegated key management. Notably, clients are not required to have expertise and infrastructure (e.g., trusted hardware modules) to properly protect keys and other secrets from being compromised. This reduces the risk of key compromise due to lack of expertise, social engineering, or other human oversight.

Indeed, clients must trust that the TEE implementation will faithfully protect the key as intended. However, significant discrepancies arise in arguments about how trust assumptions change between the two approaches, as client machines are typically equipped with similar processors as the server operating the TEE, and clients must therefore \textit{still} trust that the client processor's firmware is acting in good faith if/when handling secrets. Our approach raises the bar for attackers by delegating to the administrators the task of hardening the (server) machines carrying secrets. This opens many opportunities to improve utility and performance.

\nbf{Secure and High-Performance File Sharing.} 
The data access model in \BFS{} is similar to that of NFS, where file operations are executed at the server. We contrast this to other approaches that cache whole-files at clients and largely execute file operations at the clients~\cite{openafs,djoko_nexus_2019}. Our approach ensures that the mechanics of encrypting data for persistence are transparent to clients. In turn, this avoids having to bootstrap costly, interactive, client-to-client protocols to perform simple tasks like sharing files.

Sharing is done using typical \texttt{chmod} or \texttt{setfacl} requests. Recipients can then begin retrieving the data, encrypted under their session key, without knowledge of the persistent key. Importantly, access is asynchronously granted to recipients, without requiring to explicitly notify them or otherwise requiring them to be online during the sharing process. We contrast this with approaches that require always-online clients for sharing to effectively occur (e.g., to distribute persistent keys)~\cite{djoko_nexus_2019}, which can become prohibitive as the file system grows or access rights change frequently.

\nbf{Efficient Revocation.} Revocation in secure file systems is notoriously challenging~\cite{garrison2016practicality,ateniese2006improved}, often requiring complex protocols for generating new encryption keys, re-encrypting data under the new keys, and distributing the new keys to the clients retaining access rights. In \BFS{}, the separation between keys used to protect data for persistence and those for transport provides a dual benefit to file sharing. It revives canonical semantics of revocation: revocations are enforced through simple permission/ACL changes on files. This requires a single operation by the data owner, and revoked clients immediately lose access to the files.

\nbf{Policy Management.} Our data access model also simplifies the mechanics of other administrative tasks. Specifically, using the \BFS{} server as an escrow allows us to realize a TEE-driven reference monitor, as all requests to read or write data must pass through the \BFS{} server.
This design point resembles traditional escrow systems, but differs in that the trust in the escrow is hardware-backed, and the \BFS{} server can perform complex file system operations rather than simply key storage. The escrow therefore has three unique capabilities.

First, it can enforce access controls on data for both normal users \textit{and} administrators. For example, the \BFS{} server can allow administrators to perform compliance auditing, while user's can attest (through attestation over the server code) that the programmed policies meet reasonable expectations of user privacy (against both administrators and other users). Second, giving data visibility to the \BFS{} server enables implementing tailored optimizations server-side, such as block-level replication, prefetching, etc. Lastly, retaining a canonical POSIX API for clients decouples client and server interface dependencies---which would otherwise make it intractable to patch or implement new features server-side without incurring compatibility hazards.
To support this, the \BFS{} server exposes RPC methods for configuring file-level access controls (ACLs) and other system-wide policies.

\section{Security Analysis}
\label{sec:sec-analysis}
Below we provide an analysis of the security guarantees provided by \BFS{}, with a particular focus on confidentiality and integrity. We organize the analysis around the primary \BFS{} components---examining a concrete set of attacks against the client, untrusted host, \BFS{} server, and storage nodes. Attacks reflect those enumerated in our threat model (see~\autoref{sec:threat-model}).

\nbf{Client.} While client machines are considered trusted, an attacker who successfully compromises a client machine may obtain access to any sensitive data the client has cached locally, as well as the client's session key (and thus can temporarily impersonate them). While such is the case for any file system, clients in \BFS{} do not manage persistent secrets and therefore the attacker would not have unfettered access to file system data. We can therefore minimize the blast radius in the event of a compromise.

\nbf{Untrusted Host.}
At the untrusted host, malicious third-party software/firmware, or a co-located tenant who has gained escalated privilege, may attempt to access or corrupt messages or return codes delivered to the latter three components.
Our trusted file system core ensures that sensitive data/metadata exists in plaintext only within the TEE, while encrypted RPC messages and blocks stored outside of the TEE are opaque to the untrusted host. Our authenticated dispatch, shielded block layer, and guarded control transfer mechanisms ensure that RPC channels cannot be hijacked or replayed, and return codes cannot be manipulated to arbitrarily direct control-flow. The primary file system secret (block-encryption key) is only known to the TEE. These mechanisms ensure the untrusted host cannot tamper with file system code or data while processing client requests.

\nbf{\BFS{} Server.} While the \BFS{} server is considered trusted, in the absence of formally-verified file system code, an attacker who manages to find and exploit a weakness in the \BFS{} server code will have access to the block-encryption key and all file system data. However, the \BFS{} code must be attested by clients and therefore any deviations from a trusted state (i.e., how routines are implemented or what secrets are present) will be detected by clients. We can therefore ensure that a compromise is localized to the exploitable code, and an adversary cannot arbitrarily change the TEE functionality.

\nbf{Storage.} Storage nodes may similarly become compromised and attempt to access or tamper with blocks as they are retrieved from or written to disk. However, attacks manifesting at storage nodes are recognized and handled by the \BFS{} server as an attack by the untrusted host; our shielding mechanisms will ensure that block data read from disk is consistent with the Merkle tree, and any tampering on writes will be detected on subsequent reads.

\section{Implementation}
We implemented \BFS{} for Linux hosts in $\sim$\SI{22}{\kilo\nothing} lines of C++. It has a metadata layout similar to local \texttt{ext} file systems and is composed of client, server, and storage nodes.

\nbf{Client.} End-users or applications mount the file system to a local directory through the FUSE~\cite{vangoor_fuse_nodate} API, with file operations sent as RPC messages to the \BFS{} server. \blue{Our client implementation is thread-safe, using reader-writer locks to support entry by multiple threads.} It also supports a rich set of file operations: \texttt{getattr}, \texttt{mkdir}, \texttt{unlink}, \texttt{rmdir}, \texttt{rename}, \texttt{chmod}, \texttt{open}, \texttt{read}, \texttt{write}, \texttt{release}, \texttt{fsync}, \texttt{opendir}, \texttt{readdir}, \texttt{releasedir}, and \texttt{create}. \blue{Like NFS, the \BFS{} client also supports data and metadata caching using the client's local file system. This enables reads to quickly be served without extra network round trips, and enables writes to finish quickly and be batched and written back by a background flusher thread. The flusher thread has configurable parameters for writing back dirty data that we calibrated to match the writeback parameters/triggers for NFS.}

\nbf{\BFS{} Server/Untrusted Host.} The \BFS{} server is an ext4 implementation ported to Intel SGX. It executes file operation handlers and block management tasks like block allocation across storage nodes; our implementation supports linear and striped allocation. The server is multi-threaded, with one worker thread per client.

The untrusted host is an RPC server and client that communicates with clients and storage nodes on behalf of the \BFS{} server. We implemented a lightweight RPC library for communication between the clients, \BFS{} server, untrusted host, and storage nodes.

\nbf{Storage.} Storage nodes are simple block devices that receive block RPC requests over the network or locally. Requests are executed by memory-mapping the associated block-device file into unprotected memory (with an \texttt{mmap} syscall) and reading/writing at the appropriate block offset.

\nbf{Authentication and Access Control.} Our design primarily address access control at system-level as opposed to file-level (i.e., ensuring only clients and the \BFS{} server may access \textit{any} data in the file system). We implement file-level access control semantics similar to NFSv4 with \texttt{AUTH\_SYS}-style authentication (i.e., file read/write/execute permissions enforced on unique user IDs associated with each connected/authenticated client), but note that the \BFS{} server exposes hooks for configuring file-level and system-wide policies. We leave future work to integrating the system with mature authentication and authorization protocols such as Kerberos or OAuth.

\nbf{Encryption and Integrity.} Communication between the clients, untrusted host, \BFS{} server, and storage nodes is protected via standard AES-128 symmetric encryption. We use Galois/Counter Mode (GCM) as it protects integrity with the MAC generated as part of the encryption process. The \BFS{} server uses SGX cryptographic libraries while non-TEE \BFS{} code uses libgcrypt. While we use pre-shared keys in our implementation for simplicity, keys could be acquired through a PKI or other key-negotiation protocols~\footnote{We chose not to use standardized security protocol suites like TLS~\cite{rescorla_transport_2018} to allow us to experiment with a wide range of constructions, optimizations, and security policies in current and future work.}. Finally, as an optimization, complete mediation of encryption and integrity checking across the host-interface is ensured through the type system: functions that traverse the host-interface only accept secure types, and sensitive data must be encapsulated in these types through encryption/MAC.

\section{Performance Evaluation}
\blue{We evaluate the performance of \BFS{} under a set of micro- and macro-benchmarks drawn directly from prior works~\cite{ahmad_obliviate_2018,djoko_nexus_2019}. The workloads represent two envisioned use cases of \BFS{}: user- and application-based clients. The former may be a developer using \BFS{} as a secure personal cloud drive for documents or code, and the latter may be a company requiring secure cloud storage for an application such as a webserver.
We seek to answer the following questions:}
\begin{enumerate}
    \item \blue{\textit{How much end-to-end read/write performance can \BFS{} deliver w.r.t. state-of-the-art systems?}}
    \item \blue{\textit{How well does the \BFS{} file system layer perform w.r.t. state-of-the-art SGX-based (local) file systems?}}
    \item \blue{\textit{How well does \BFS{} perform on complex workloads involving a mix of read/write and metadata operations?}}
\end{enumerate}

\subsection{Experiment Setup}
\nbf{Testbed.} Similar to other works~\cite{shinde_panoply_2017,djoko_nexus_2019}, clients/server run on a local cluster containing Intel Core i7-10710U \SI{1.10}{\giga\hertz} processors with 12 logical cores, \SI{32}{\giga\byte} memory, and locally attached Samsung 980 Pro NVMe SSDs. All machines are Debian-based and connected in a local network over \SI{1}{\giga bE} interfaces. The SGX code is compiled in HW mode using SDK version 2.24, the native SGX driver in the Linux 5.15 kernel, and the standard 128 MB EPC. The non-TEE code uses libgcrypt v1.8.5.

\nbf{Baselines.} 
\label{sec:baselines}
\blue{\ifrevision\hypertargetblue{r32}{\re[2]{3}}~\fi We compare the performance of \BFS{} against several other systems. First, we compare against a non-SGX version of \BFS{} (with SGX ocalls and ecalls simply replaced by direct function calls) \blue{to measure SGX overheads}. Next, we compare against \textbf{NFS} (with/without Kerberos network encryption), because it is the industry-standard for cloud file system deployment (e.g., used in AWS EFS and Google Filestore~\cite{haynes_network_2015,aws-efs,google-filestore}). We compare against \textbf{NeXUS}~\cite{djoko_nexus_2019}, an SGX-based cloud file system that handles all tasks client-side, using the cloud as a simple persistent storage. We note that NeXUS appears only in half of the micro-benchmarks and no macro-benchmarks, because the code does not support direct I/O and simply crashes when trying to run any non-trivial workload. We also compare against \textbf{Gramine}~\cite{tsai_graphene-sgx_nodate,gramine-proj} (formerly Graphene-SGX), because it is decidedly the state-of-the-art in SGX-based file systems, being backed by several industry partners and under active development by the Linux Foundation~\cite{gramine-ccc}. Finally, we compare against \textbf{ZeroTrace}~\cite{sasy_zerotrace_2018}, which is another SGX-based local storage interface that provides oblivious access to generic array data structures (e.g., an array of file data blocks) with ORAM techniques. Though \BFS{} provides a more comprehensive set of security guarantees than the latter two systems, we compare against them to understand the viability of \BFS{} in practice.}

The \BFS{} and NFS system configurations provide different confidentiality and integrity guarantees and are denoted as follows: (1) \texttt{nfs\_ne}, the NFS baseline without any encryption; (2) \texttt{nfs\_we}, NFS with Kerberos-based network encryption and integrity protection (i.e., mounted with \textit{sec=krb5p}) but no disk encryption; (3) \texttt{bfs\_ne}, the \BFS{} baseline without SGX memory encryption; and (4) \texttt{bfs}, \BFS{} with full encryption and integrity protection (network, memory, and disk). Importantly, we note that \texttt{bfs} provides additional security guarantees (memory and disk encryption) over \texttt{nfs\_we}, the NFS deployment mode typically used in practice. 

\begin{figure*}[!t]
    \centering
    \includegraphics[width=0.49\textwidth]{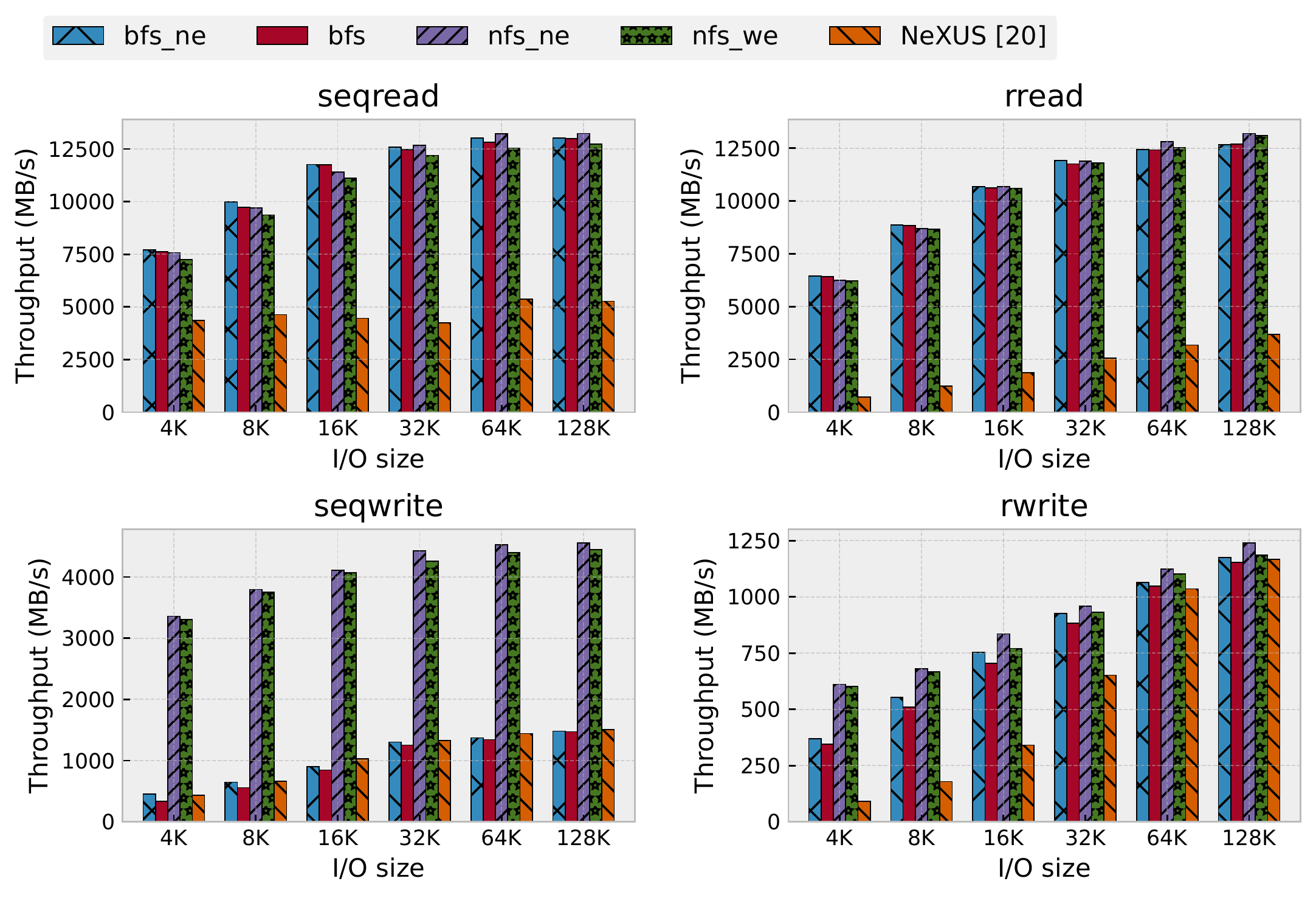} %
    \includegraphics[width=0.49\textwidth]{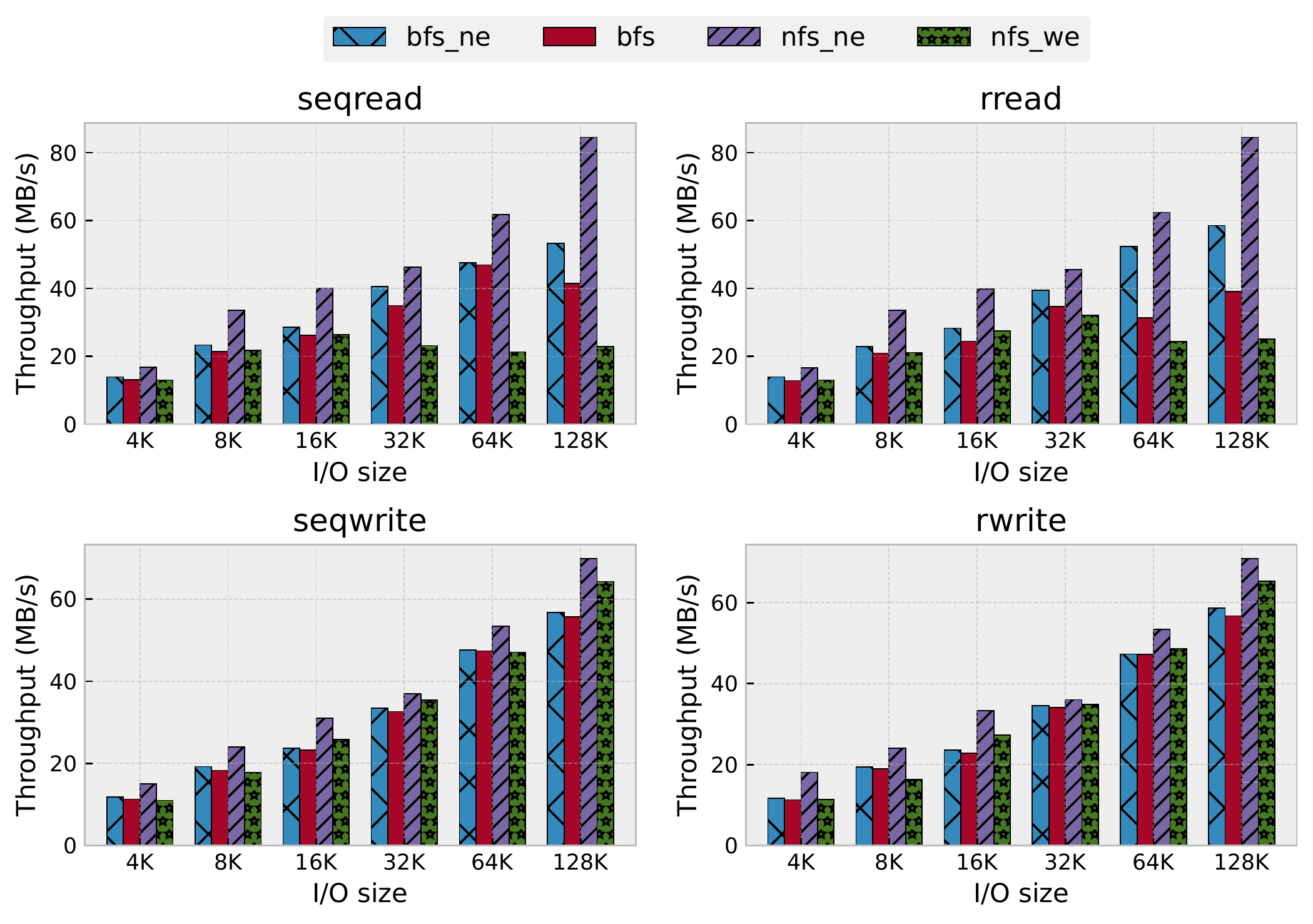} %
    \caption{Micro-benchmark performance. Single-client, local storage, with client caching (left) and without client caching (right).}
    \label{fig:bench-micro-local}
\end{figure*}

\nbf{Client Optimizations.} \blue{Client optimizations are a central component that enables NFS to minimize RPC calls to the server and support high-performance applications. Of note are NFS's use of client-side data and metadata caches (which enable fast read/write paths), compound operations (which batch multiple RPCs to reduce round trips), and delegations (which enable clients to act autonomously when performing certain operations like opening/closing files)~\cite{haynes_network_2015}. While our implementation currently supports client-side data and metadata caches, the core innovation of \BFS{} is not client design. In addition to running workloads with client caching enabled, we therefore also attempt to isolate the effects of sophisticated NFS client optimizations to accurately measure the raw performance that the \BFS{} server can deliver w.r.t. the NFS server. We do this by running \textit{direct I/O}-based workloads with the \textit{noac} NFS mount option (similarly toggled in \BFS{}), which partially isolates optimizations. Note that there is no straightforward way to disable other NFS client optimizations. Other mount options follow best-practice~\cite{aws-mo}.}

\subsection{Micro-benchmarks}
\label{sec:micro}
\blue{We first study the performance of \BFS{} under standard micro-benchmarks~\cite{ahmad_obliviate_2018,djoko_nexus_2019,kannan_designing_nodate}. We aim to understand the SGX overhead costs and the raw achievable read/write performance of \BFS{} at various I/O sizes. The workload profiles are provided by Filebench~\cite{Tarasov2016FilebenchAF} and are single-threaded; we will analyze multi-threaded performance in the macro-benchmarks. The seqread workload performs sequential reads of the specified I/O size on a preallocated 1GB file. The rread workload performs random reads of the specified I/O size on a preallocated 1GB file. The seqwrite workload writes a new 1GB file sequentially with the specified I/O size. The random write workload performs random overwrites of the specified I/O size on a preallocated 1GB file. All benchmarks run for 10 minutes to allow a large number of I/Os to complete, and the mean throughput across 10 independent trials is taken. Caches are flushed between each experiment.}

\blue{\autoref{fig:bench-micro-local} shows the performance under each workload, with client caching enabled in the four graphs on the left and disabled in the four graphs on the right. The general trend across all graphs is that throughput increases with I/O size. This indicates that clients are able to make more efficient use of link bandwidth per-request by using larger I/Os (i.e., RPC messaging costs are amortized).}

\nbf{\blue{SGX Overheads.}}
\blue{As \texttt{bfs\_ne} runs without SGX enabled (the same code, with direct function calls used in place of SGX call gates), it serves as our baseline for understanding the relative overhead of using the TEE. With client caching enabled, we observe that on average \texttt{bfs} delivers $>90\%$ of the throughput of \texttt{bfs\_ne}. In fact, across all workloads, even at large 128K I/Os, \texttt{bfs} nearly matches the performance of \texttt{bfs\_ne}. We attribute this to the client cache being able to largely absorb overheads associated with disk encryption and the Merkle tree by handling read and write requests client-side.}

\blue{With client caching disabled, we observe that \texttt{bfs} can deliver similar write performance to \texttt{bfs\_ne}. However, the read performance difference is more significant (approx. 50\%) at large I/O sizes. We attribute this to SGX paging overheads, which can affect certain workloads in unpredictable ways~\cite{taassori2018vault}. We will revisit this point below.}

\nbf{\blue{Comparison to NFS~\cite{haynes_network_2015}/NeXUS~\cite{djoko_nexus_2019}}.}
\blue{With client caching enabled, we observe that \texttt{bfs} delivers nearly the same read throughput as \texttt{nfs\_ne} and \texttt{nfs\_we}, up to $12.5$~GB/s with 128K I/Os. We note that these throughputs reflect those seen in practice under the same mount options, which can range from a few hundred MB/s to several GB/s~\cite{aws-efs-perf}. \texttt{bfs} also substantially outperforms NeXUS, up to $2.5\times$ with 128K I/Os.  We attribute this to \BFS{} having a more efficient client design than NeXUS, and the client cache being as efficient as the NFS client cache in general---which allows clients to handle reads at a speeds approaching DRAM speed.
We note that the \BFS{} client design is fundamentally different than NeXUS's. In NeXUS, clients perform all cryptographic work, and this approach has many practical limitations (\autoref{sec:key-mgmt}). Notably, NeXUS does not prevent rollback attacks (as noted by the authors), and it requires clients to store the entire file system locally, which is intractable at large-scale. Further, the client-to-client nature of file sharing necessitates always-online clients.}

\blue{With client caching disabled, we observe that \texttt{bfs} delivers up to a $2.2\times$ speedup over \texttt{nfs\_we} on read performance. We attribute this to our unified, user-space server design, whereas the NFS server runs in the kernel and often requires expensive upcalls to user-space daemons~\cite{haynes_network_2015}.  However, as similarly observed above against \texttt{bfs\_ne}, the read performance difference with \texttt{nfs\_ne} is more significant at large I/O sizes. Yet, read caching is a standard optimization for many applications---web servers in particular, which are a common TEE-based application. And client caching can largely mask read overheads and enable \texttt{bfs} to deliver near-equivalent end-to-end user/application performance to \texttt{bfs\_ne} and \texttt{nfs\_ne}. Thus, from a practical standpoint, the performance difference here is not a fundamental problem.}

\blue{With client caching enabled, \texttt{bfs} delivers similar random write performance to \texttt{nfs\_ne} and \texttt{nfs\_we} at I/O sizes larger than 4K, with up to a $4\times$ speedup over NeXUS at smaller I/O sizes. However, the sequential write performance difference with \texttt{nfs\_ne} and \texttt{nfs\_we} is significant. Sequential writes are write-allocating, meaning that they cannot exploit data caches. We found that the NFS client optimizations (compound operations and delegations) more efficiently handle all of the internal RPC calls associated with write-allocating \texttt{write()} syscalls. We therefore attribute this performance difference to \BFS{} having a less efficient \textit{client} design than NFS, particularly in the context of handling sequential writes.}

\blue{However, as seen from the results with client caching disabled, \texttt{bfs} delivers $>85\%$ of the throughput of \texttt{nfs\_ne} and \texttt{nfs\_we} for sequential writes. This implies that the \BFS{} server is nearly as efficient as the NFS server at handling writes. As mentioned, the core innovation of \BFS{} is not client design. Thus, we conclude that the performance difference with client caching is not fundamental to the \BFS{} server design, as further client improvements can help eliminate overheads~\footnote{\blue{To enable better integration into real systems, we are currently extending the NFS server implementation to enable callbacks into the \BFS{} server code, such that \BFS{} exports can be exposed transparently to standard NFS clients.}}.}

\begin{figure}[!t]
    \centering
    \includegraphics[width=0.5\textwidth]{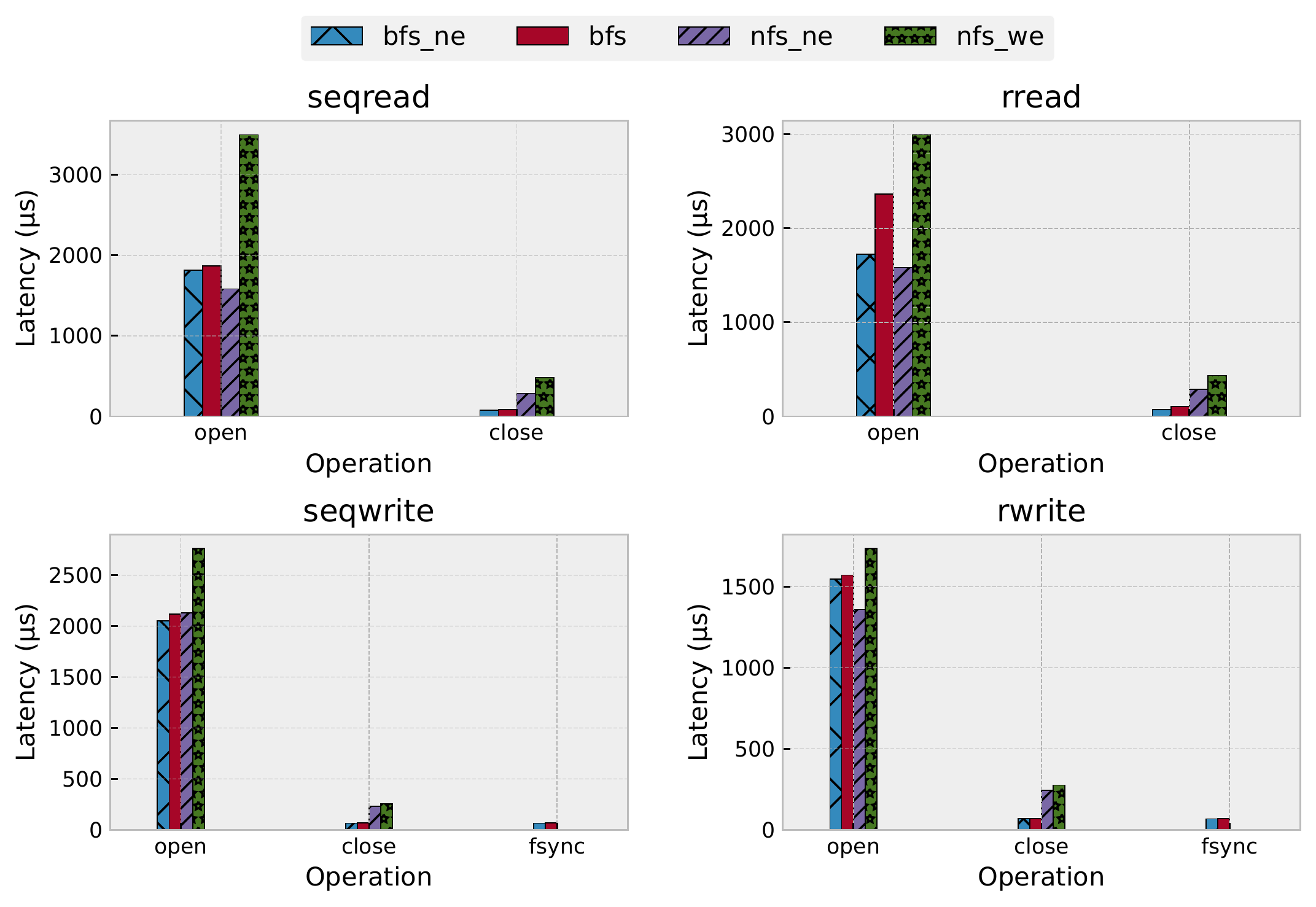}
    \caption{\BFS{} shows practical performance on metadata operations. \texttt{bfs} has lower open and close latencies than \texttt{nfs\_we} across all workloads. \texttt{bfs} has higher fsync latency, but the root cause is NFS client optimizations which delay actual fsyncs until subsequent closes/opens.}
    \label{fig:meta-perf}
\end{figure}

\nbf{\blue{Metadata Operations.}}
\blue{In \autoref{fig:meta-perf}, we examine the latencies of performing critical metadata operations: open, close, and fsync (to force file data buffered at the server to disk). Latencies are averaged across I/O sizes shown in~\autoref{fig:bench-micro-local}. Note that NFS client optimizations like delegations and compound operations have more significant and unpredictable effects on client performance with client caching enabled, which can skew measurements of the underlying server performance. We therefore focus on the case without client caching to isolate NFS client optimizations as much as possible (which \BFS{} does not implement) for a fairer comparison.}

\blue{\autoref{fig:meta-perf} shows that \texttt{bfs} has lower open and close latencies than \texttt{nfs\_we} across all workloads. In particular, \texttt{bfs} has nearly half the latency of \texttt{nfs\_we} on sequential reads. Interestingly, we also observe that \texttt{bfs} has higher fsync latency than \texttt{nfs\_we} and \texttt{nfs\_ne}. However, NFS allows clients to delay COMMIT RPCs generated by fsync calls until subsequent file closes or opens. This is why fsync appears extremely fast ($<1\mu s$) and why NFS opens and closes appear relatively slower than \BFS{}.}

\blue{We do not conclude that \BFS{} can necessarily open or close files faster than NFS, nor that \BFS{} fsync is slower. There is a delicate trade-off between deciding when to flush data and the user/application-perceived performance. Delaying actual flushes (as NFS does to ensure close-to-open consistency) can allow an application to proceed with more work in real-time but weakens durability guarantees. Not delaying flushes can ensure stronger write durability but at the expense of users/applications possibly encountering latency spikes when fsync is called (e.g., whenever a developer saves a source file in their editor). However, although \texttt{bfs} fsync latency is higher than NFS, the latency across both write workloads is $<73\mu s$. This is still far below latency thresholds deemed to be perceived by users as instantaneous ($<100ms$)~\cite{miller1968response}. Further, for applications like databases, write stalls may occur frequently if fsync cannot keep up with incoming write requests. However, the mean fsync latency of $<73\mu s$ we observed across I/O sizes for the seqwrite/rwrite workloads for \texttt{bfs} is significantly less than the mean write latency of $<1039\mu s$. We also saw in~\autoref{fig:bench-micro-local} that \texttt{bfs} already delivers write performance on par with NFS. Thus, we conclude that \BFS{} write performance is not fundamentally bottlenecked by this higher fsync latency.}

\begin{figure}[!t]
    \centering
    \includegraphics[width=0.5\textwidth]{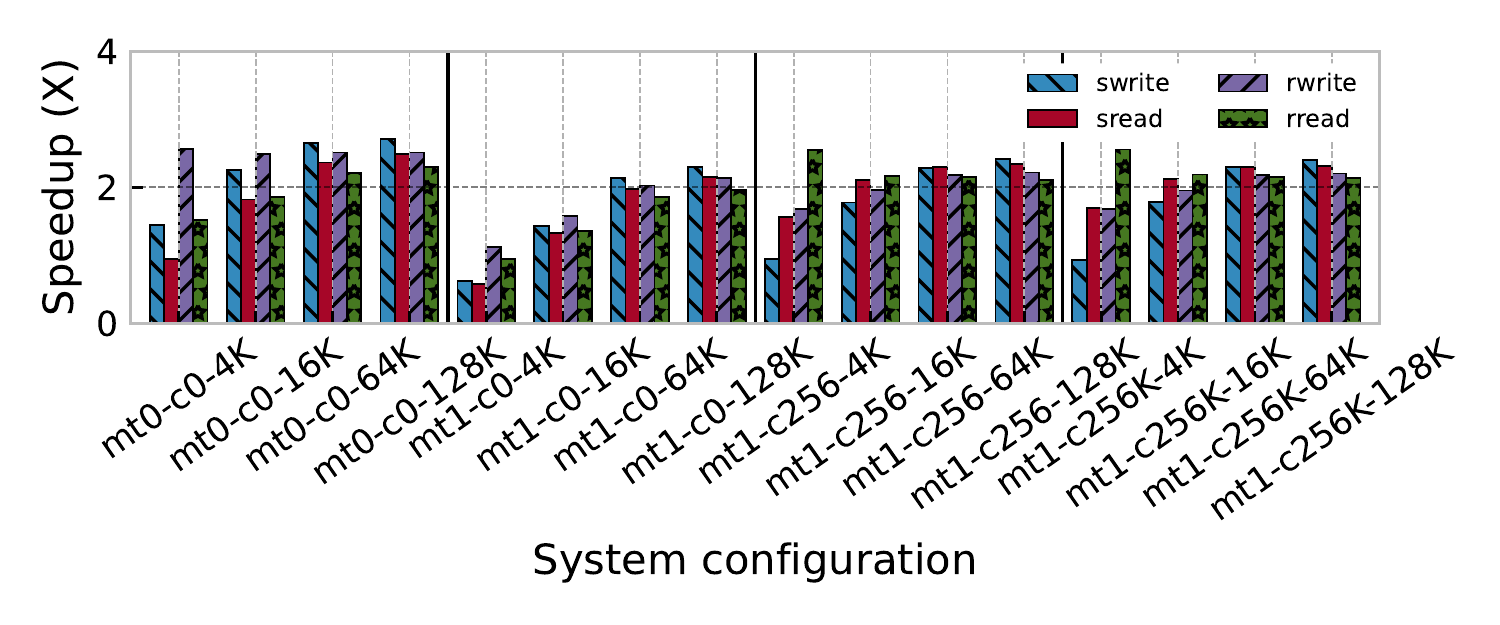}
    \caption{Speedup of \BFS{} over Gramine~\cite{tsai_graphene-sgx_nodate,gramine-proj}. Configuration parameters are encoded in the form \textit{a-b-c}, where \textit{a} denotes whether the Merkle tree is enabled (1) or not (0), \textit{b} denotes what cache size is used, and \textit{c} denotes the I/O size used.}
    \label{fig:gramine-speedup}
\end{figure}

\nbf{\blue{Comparison to Gramine~\cite{tsai_graphene-sgx_nodate,gramine-proj}.}} \blue{Next, we run the same workloads directly against the Gramine file system implementation. Note that there is no frontend client for Gramine available to measure end-to-end user/application performance. Our goal is therefore to measure backend performance. Specifically, we want to measure the trade-offs between the two approaches to file system design discussed above. \autoref{fig:gramine-speedup} shows several different configurations, parameterized by workload, whether the Merkle tree is enabled, whether \BFS{} uses an in-TEE block cache, and what I/O size is used.}

\blue{\BFS{} delivers $1-2.5\times$ speedups over Gramine. The largest speedups are observed under three configurations: at large I/O sizes, when the Merkle tree is enabled, or when \BFS{} maintains a small block cache (256 blocks = 1MB). We attribute these speedups to \BFS{} having less software layers between the read/write call entry point and storage, whereas Gramine has several added layers of abstraction.} \blue{We note that Gramine always caches entire files in the TEE, verifies the hashes only once (when the file is opened), and flushes them out of the TEE only once (when the file is closed). This can quickly become prohibitive for large files. In contrast, \BFS{} verifies hashes and flushes hash updates on every file read/write, and still shows performance improvements.}

\blue{This result is significant because there has been a significant push in the community to \textit{minimize} the amount of code running in the TEE, for fear of severe performance loss due to SGX paging (and context switching)  overheads~\cite{taassori2018vault,orenbach_eleos_2017}. The result is that the Gramine libOS model reigns supreme, where a thin wrapper library runs in the TEE and the untrusted host executes most of the server logic. Yet, the libOS model exposes large and complex host-interfaces, which are hard to shield in general---and state-of-the-art systems like Gramine have limited shielding mechanisms in practice. This invites unnecessary security risks~\cite{van2019tale}. \BFS{} instead shows that it is possible to deliver practical performance with \textit{more} logic running in the TEE and a minimal host-interface.}

\nbf{\blue{Comparison to ZeroTrace~\cite{sasy_zerotrace_2018}.}}
\blue{Next, we run the random read/write workloads against ZeroTrace. ZeroTrace is a library that can be statically linked into an SGX-based application, and it provides oblivious access to array data structures---in our case, this is an array of storage data blocks. ZeroTrace also encrypts data and uses a Merkle tree to ensure freshness. Again, there is no frontend client to measure end-to-end performance, so we focus on backend performance. Further, as noted by the authors in the code repo, the code for on-disk data structures is broken, so we report ZeroTrace results on an in-memory array, which underestimates the \BFS{} speedups reported below. We use the default ORAM parameters the authors used in their benchmark scripts.}

\begin{figure}[!t]
    \centering
    \includegraphics[width=0.45\textwidth]{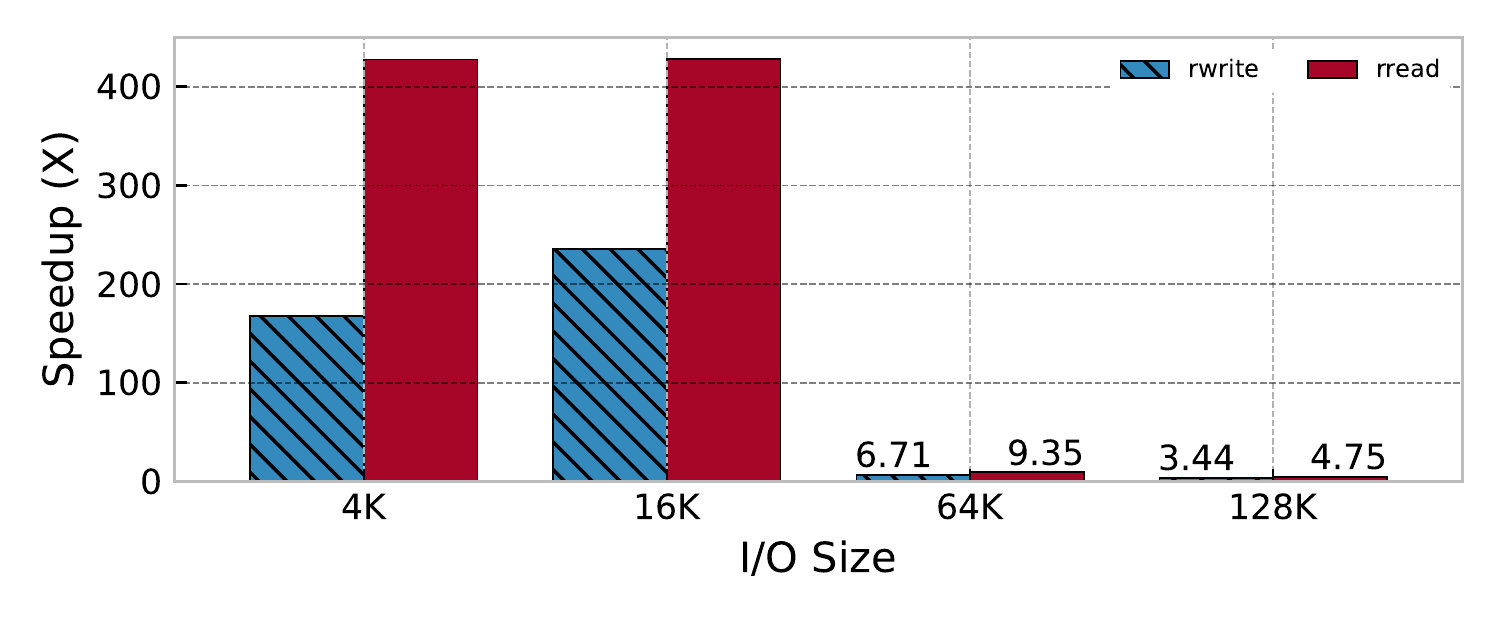} %
    \caption{Speedup of \BFS{} over ZeroTrace~\cite{sasy_zerotrace_2018}. \BFS{} can deliver better performance at small and large I/O sizes.}
    \label{fig:zt-speedup}
\end{figure}

\blue{\autoref{fig:zt-speedup} shows the speedup of \BFS{} over ZeroTrace at various I/O sizes. First, we observe that \BFS{} has nearly a $200\times$ and $410\times$ speedup over ZeroTrace on 4K random write and read performance, respectively. Initially, increasing the I/O size only worsened ZeroTrace's performance, because it required additional (costly) ORAM fetches (which are normally at 4K granularity, our block size). When examining larger I/O sizes, we therefore instead increased ZeroTrace's $block\_size$ parameter to reduce per-fetch costs. We still observed that \BFS{} had at least a $3.44\times$ and $4.75\times$ write and read speedup, respectively, at 128K I/Os. \BFS{} will likely see larger speedups when using larger block sizes.}

\blue{\ifrevision\hypertargetblue{r12}{\re[2]{1}}~\fi As a local storage interface, this shows that oblivious interfaces still have a long way to practicality. The core innovation of \BFS{} is that it lifts the core file system layer into the TEE to ensure all data and metadata are isolated from the untrusted host, exposing a minimal block-level host-interface. Thus, from a practical standpoint, \BFS{} raises the bar for attackers as high as possible without resorting to ORAM.}

\nbf{\blue{\ding{228}~Takeaway:}}
\blue{These results lead us to two key conclusions. 1) SGX costs can largely be absorbed with client caching, enabling performance close to state-of-the-art cloud file systems like NFS. 2) \BFS{} offers better performance and practical advantages compared to NeXUS, stronger security guarantees than Gramine, and balances the security-performance trade-off more efficiently than ZeroTrace.}

\subsection{Macro-benchmarks}
\label{sec:macro}
\blue{Next, we study the performance of \BFS{} across various macro-benchmarks exercising more complex mixes of read/write and metadata operations. Like prior works~\cite{djoko_nexus_2019}, our workloads include standard Linux utilities (to emulate user-based clients) and various Filebench workload profiles (to emulate application-based clients). The utility workloads are single-threaded, and the Filebench workloads are multi-threaded (with up to 200 reader/writer threads). Filebench workloads run for 10 minutes, and mean throughput across 10 independent trials is taken. The utilities run until completion, and mean latency across 100 trials is taken. Caches are flushed between each workload run.}

\begin{figure}[!t]
    \centering
    \includegraphics[width=0.45\textwidth]{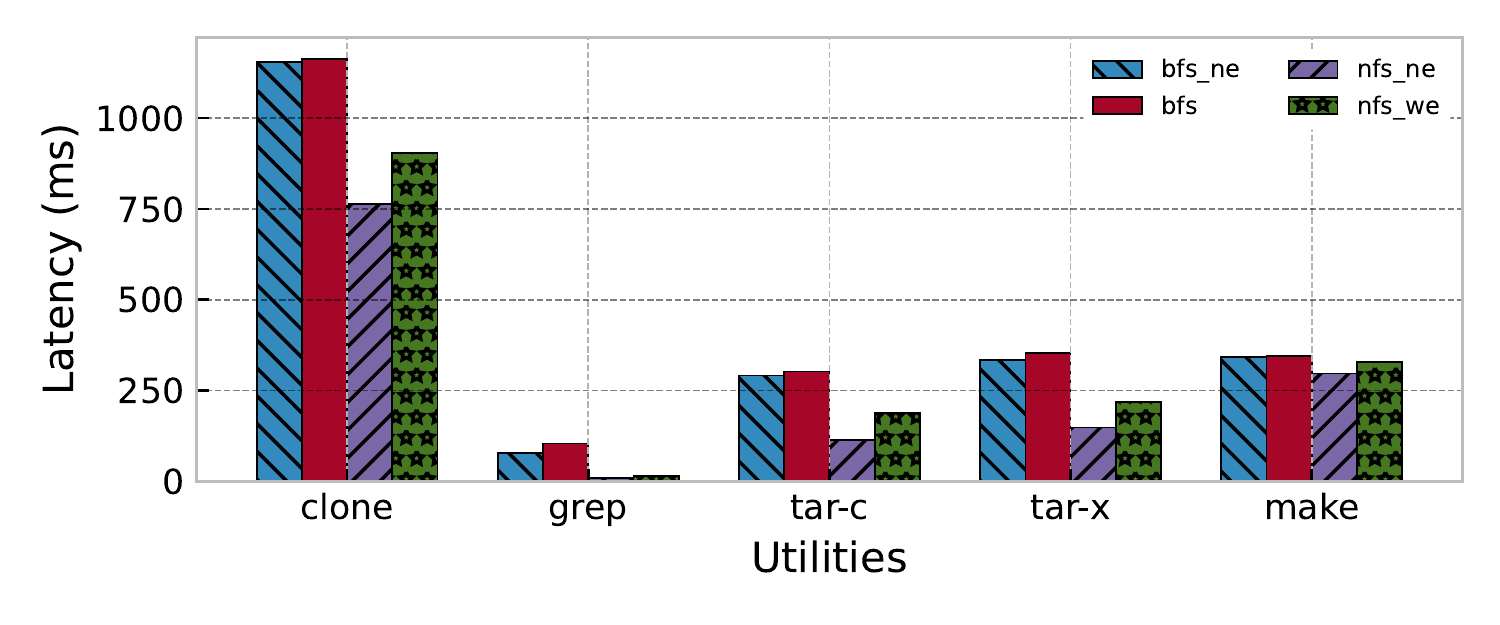}
    \includegraphics[width=0.45\textwidth]{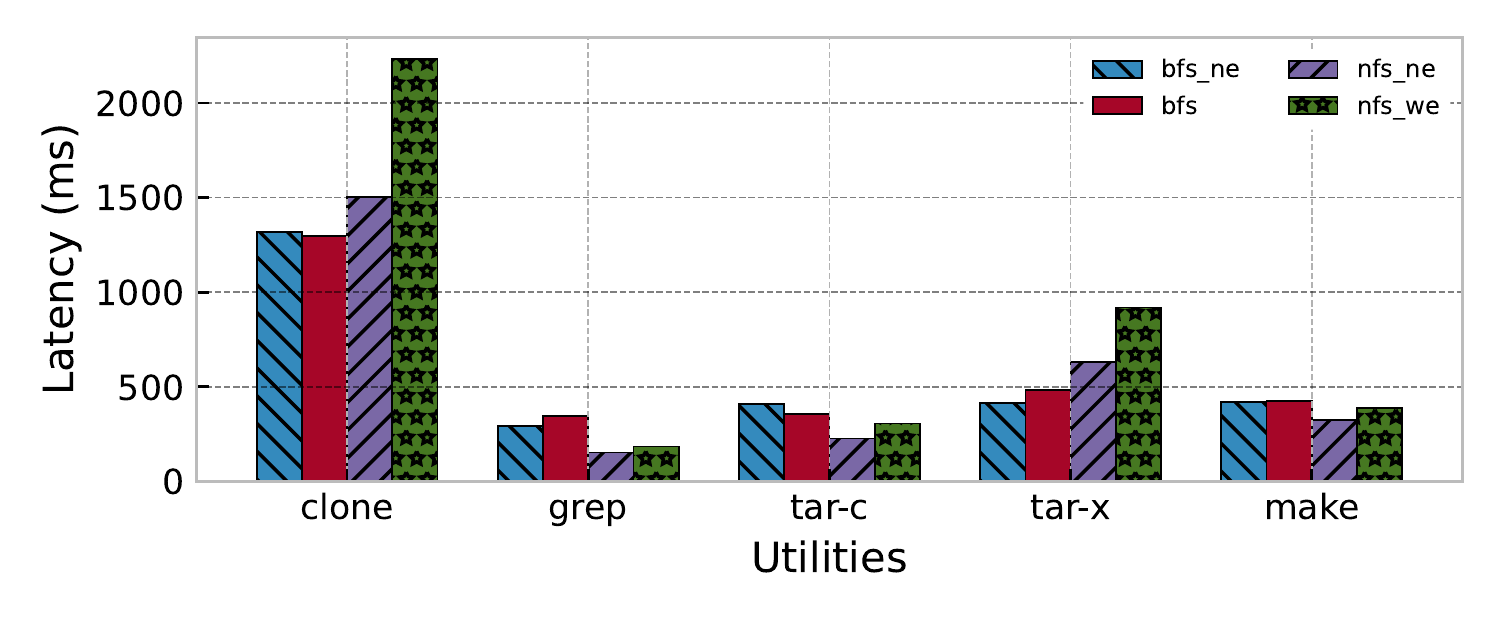}
    \caption{Task latency of various Linux utilities, with client caching (top) and without (bottom).}
    \label{fig:utilities}
\end{figure}

\nbf{\blue{User-based Clients.}}
\blue{Analyzing Linux utilities helps understand how user-based clients will perceive the performance of their networked storage~\cite{djoko_nexus_2019}. The five workloads include: \texttt{git-clone}, which clones the public linux-sgx-driver repo to a folder on the \BFS{}/NFS mount (containing approx. 20 files up to 25K in size), \texttt{grep}, which searches for 100 random strings in the repo, \texttt{tar-c}, which creates a new tar archive from the repo, \texttt{tar-x}, which extracts the contents of the tarred repo, and \texttt{make}, which compiles the driver code.} \blue{\autoref{fig:utilities} shows the performance across each utility, with/without client caching.}

\blue{As observed in the micro-benchmarks, \texttt{bfs} has near-equivalent performance to \texttt{bfs\_ne} across each workload and with/without client caching. With client caching, \texttt{bfs} sees $<1\times$ overhead over \texttt{nfs\_we} across all workloads except for grep. We find that grep exercises many \texttt{stat} system calls, which get translated into several RPCs to the server, which NFS can efficiently batch together while \BFS{} cannot. In fact, without client caching, \texttt{bfs} has $<1\times$ overhead over \texttt{nfs\_we} across \textit{all} benchmarks, and in fact sees \textit{lower} latency than both \texttt{nfs\_ne} and \texttt{nfs\_we} on the clone and tar-x benchmarks.}

\blue{We reason that, without client caching, the \BFS{} client can perform certain metadata operations quicker than NFS, and in aggregate, perform better on certain workloads. With client caching, other NFS optimizations become more effective.
Nonetheless, this shows that \BFS{} provides reasonable overheads for user-based clients. By extending the NFS server implementation to hook into \BFS{} callbacks, \BFS{} clients will be able to take advantage of all NFS optimizations and \BFS{} will approach the full speed of NFS.}

\begin{figure}[!t]
    \centering
    \includegraphics[width=0.4\textwidth]{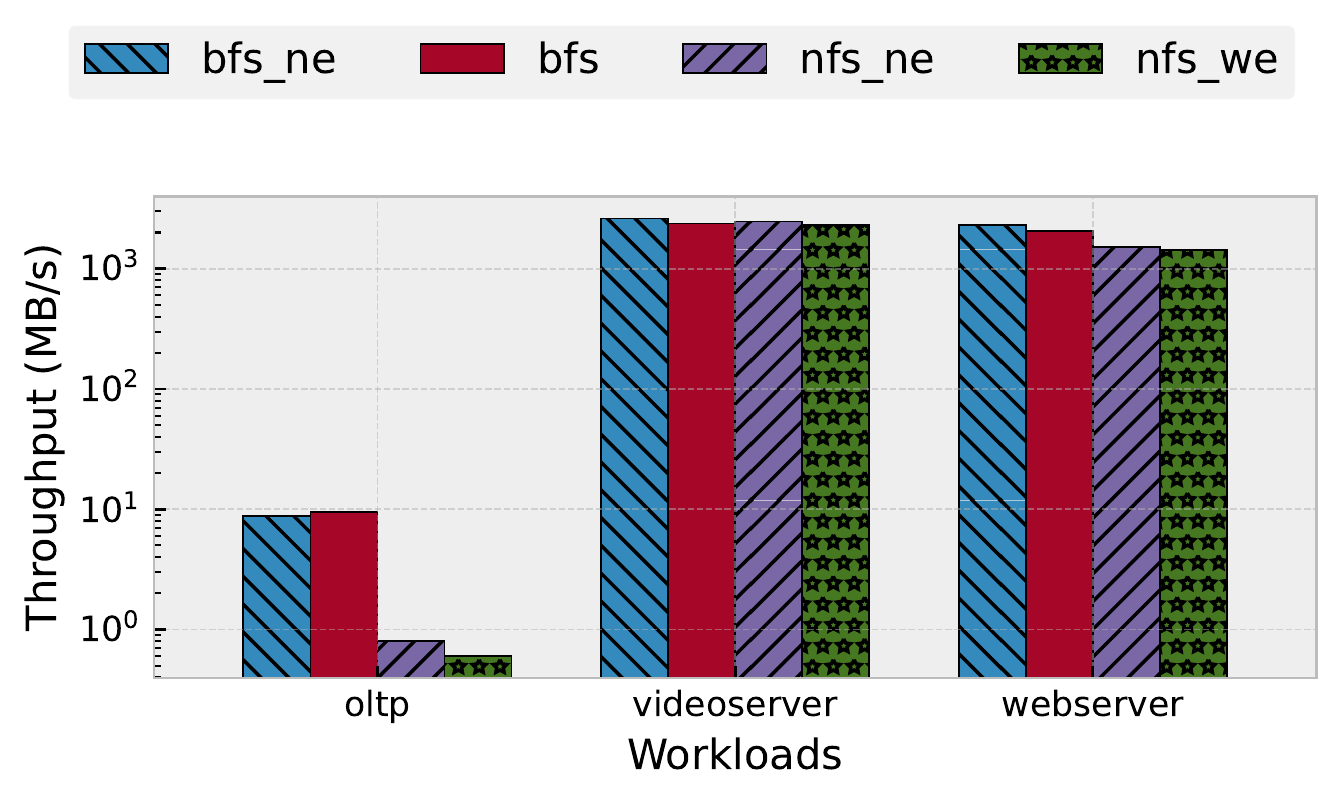}
    \includegraphics[width=0.4\textwidth]{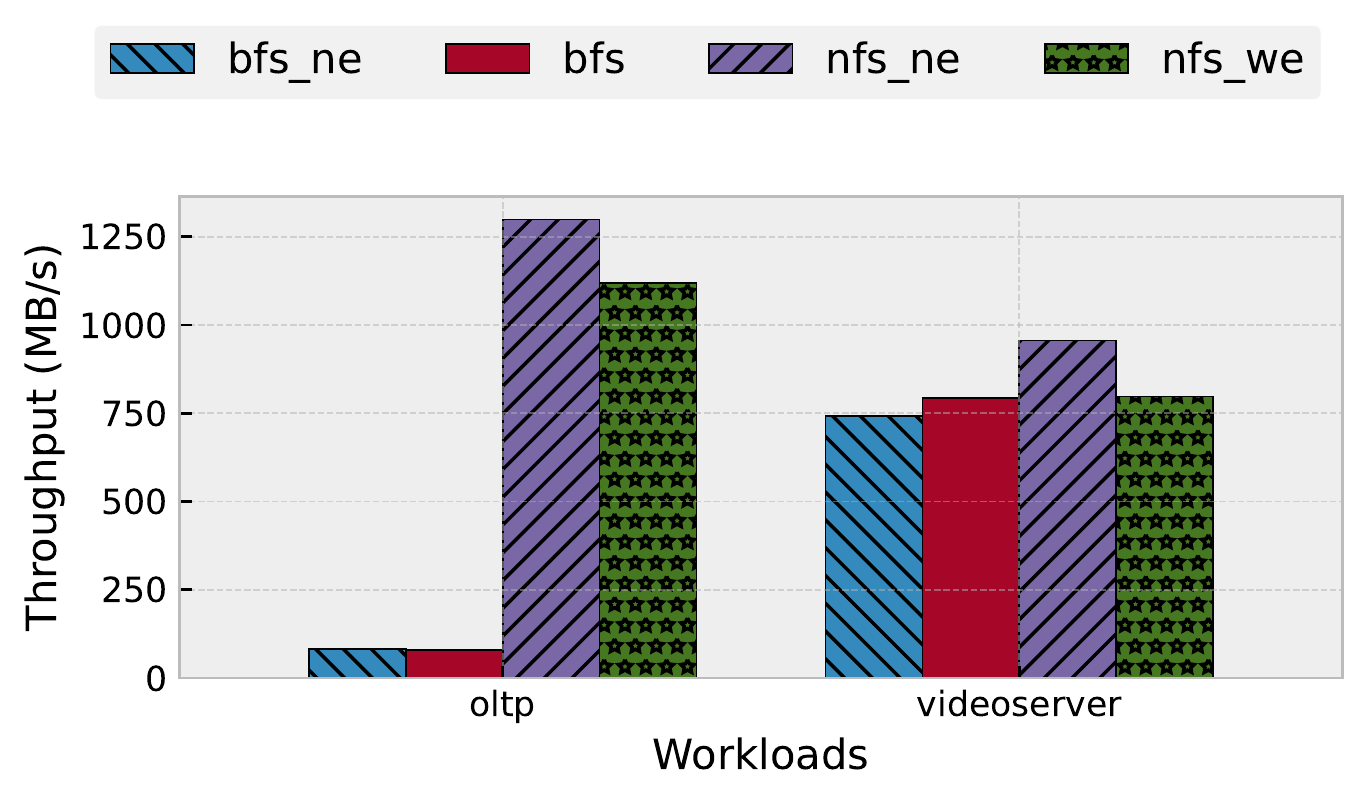}
    \caption{Filebench macrobenchmark workloads, with client caching. Read (top) and write (bottom) performance.}
    \label{fig:fb-macro}
\end{figure}

\nbf{\blue{Application-based Clients.}}
\blue{Next we focus on real-world applications that are commonly run within a TEE~\cite{tsai_graphene-sgx_nodate,arnautov_scone_nodate}. The Filebench workloads include: \texttt{oltp}, which emulates an online transaction processing system with 10 writer threads, 200 reader threads, and 100 files totaling approx. 10G in size; \texttt{videoserver}, which emulates a video server with 1 writer thread, 48 reader threads, and 226 files totaling approx. 226G in size; and \texttt{webserver}, which emulates a webserver with 100 reader threads and 10K files totaling approx. 1G in size.}

\blue{We first observe that \texttt{bfs} has near-equivalent performance to \texttt{bfs\_ne} across each workload; client caching largely absorbs SGX overhead costs. Interestingly, we also observe that \texttt{bfs} delivers nearly $14\times$ more read throughput than \texttt{nfs\_we}, but $14\times$ less write throughput, on \texttt{oltp}. We do not conclude that \BFS{} necessarily delivers higher read throughput than NFS, nor less write throughput. The NFS client also has several other optimizations for prioritizing RPC requests (and internal lock requests) initiated by certain syscalls. \BFS{} does not prioritize requests, and since \texttt{oltp} has 200 readers, reader threads tend to acquire and hold read locks more frequently. Nonetheless, \texttt{bfs} delivers nearly the same read/write throughput \texttt{nfs\_we} on \texttt{videoserver} and \texttt{webserver}.}

\blue{This shows that \BFS{} can deliver high performance under high concurrency, and more complex workloads that exercise more metadata operations with read/write operations.}

\nbf{\blue{\ding{228}~Takeaway:}}
\blue{These macro-benchmark results further support the efficacy of the \BFS{} design. \BFS{} largely follows the NFS model, and as such, can deliver high-performance on many workloads. And compared to state-of-the-art systems, we show that many concerns raised regarding SGX overheads can largely be mitigated with judicious client design.}

\section{Discussion}
\label{sec:discussion}

Below we discuss notable points of consideration for \BFS{}.

\nbf{Extending to Other TEEs.}
A core security guarantee of any TEE implementation is that it provides a notion of isolation between trusted and untrusted code running on a shared machine. For example, {ARM TrustZone} TEEs are characterized by a secure and non-secure ``world'' or state, where all memory has an extra bit defining its state~\cite{arm_security_nodated}. {AMD SEV} implements a VM based TEE, providing separation at the boundary of the secure VM.

\blue{\ifrevision\hypertargetblue{r21}{\re[1]{2}}~\fi While our implementation is based on SGX, the design of \BFS{} is not fundamentally tied to SGX. The \BFS{} server only needs a mechanism to send/receive incoming network messages from clients on the frontend and send/receive incoming messages to storage devices on the backend. There are many SDKs and third-party libraries available to enable this across a wide variety of TEE implementations~\cite{openenclave}. The \BFS{} server can therefore be ported to any TEE platform. Note that the security guarantees will only be as good as those afforded by the TEE implementation (which may vary between vendors).} We leave future work to extending \BFS{} to other TEEs.

\nbf{Limitations.} \label{sec:discussion:limitations} While our design successfully defends against a wide range of known and new attacks, side-channels attacks on TEEs still present a challenge~\cite{gotzfried_cache_2017-1,lee_inferring_nodate,oleksenko_protecting_nodate,nilsson_survey_2020}. \blue{Most work studying oblivious access mechanisms to defend against side-channel attacks have focused on exploiting rich interfaces, such as database queries. While here the untrusted host can see block addresses in RPC messages, it remains an open question to what extent low-level block access patterns can be traced back to file system data or application logic. We discussed how ZeroTrace can be used to defend against potential side-channel attacks, but not without making a steep performance trade-off. We leave future work to exploring side-channel attacks and efficient mitigations at the block layer in more depth.}

\nbf{Other Backend Architectures.}
Our current \BFS{} design focuses on a single-server scenario and centralized storage management (i.e., both data and metadata are managed at the same server). \blue{\ifrevision\hypertargetblue{r31}{\re[1]{3}}~\fi This architecture is emblematic of widely popular cloud file systems offered by major cloud providers (e.g., AWS EFS, Google Cloud Filestore), and is the file system of choice for many cloud applications. We therefore focused our analysis on these classes of workloads and focused analysis against NFS. Other file systems like Ceph have taken a decentralized approach to metadata management to improve performance, largely for HPC environments~\cite{weil_ceph_nodate}. \BFS{} could be extended to a decentralized architecture like Ceph. We leave such a design and implementation to future work.}

\section{Related Work}
\label{sec:related}

Cloud file system design has a long history that intersects storage, applied cryptography, and trusted computing research. \BFS{} builds on the lessons learned in these works, rethinking the fundamental file system abstractions to produce a design with a unique set of capabilities.

\nbf{Client-based Solutions.} Client-based solutions have been the standard approach to designing secure outsourced storage systems. For example, CFS~\cite{blaze_cryptographic_1993}, Plutus~\cite{kallahalla_plutus_2003}, and NeXUS~\cite{djoko_nexus_2019} require clients (or trusted client proxies) to execute file operations and handle all cryptographic tasks, while files are organized as opaque blobs on the untrusted server/storage. Other works also assume a client-based gateway to untrusted storage~\cite{stefanov_iris_2012,chen_seminas_2016,vrable2012bluesky,viotti2017hybris}. While perhaps useful in some contexts, such designs are ill-fit for typical usage patterns of cloud storage; \blue{NFS is still decidedly the state-of-the-art cloud file system used in practice. First, such approaches typically do not ensure rollback protection~\cite{djoko_nexus_2019}, as it requires costly client-to-client coordination on every update to file data. Second, they require clients to store the entire file system locally, which is intractable at large-scale. Further, they burden clients with having to execute complex protocols to perform simple tasks like file sharing, and they require clients to have proper training and expertise with managing keys. Clients commonly expect a POSIX-like interface with similar guarantees to NFS (close-to-open consistency), and key management is most often delegated server-side.} Running a common application such as collaborative document editing is infeasible if not practically impossible on top of such systems.

In \BFS{}, we instead delegate encryption for persistence to the file server (in our design, denoted as the \BFS{} server) and revive the NFS model by redesigning the structure of the file server to provide stronger security guarantees (against more attacks), extensible feature support, and practical performance.

\nbf{LibOS Runtimes.} Many recent efforts have relied on libOS runtimes as a means for quickly porting server applications to use TEEs. LibOSes enable applications to run unmodified in TEEs by automatically generating the necessary wrapper code (including encryption/decryption operations over data) to redirect system calls from within the TEE onto the host~\cite{baumann_shielding_2015,arnautov_scone_nodate,tsai_graphene-sgx_nodate,shinde_panoply_2017,priebe_sgx-lkl_2020,thalheim_rkt-io_2021}. Typical storage applications include in-memory databases~\cite{priebe_enclavedb_2018}, local file systems~\cite{ahmad_obliviate_2018}, and key-value stores~\cite{bailleu_speicher_nodate,bailleu_avocado_nodate}. However, simply porting a traditional file system (like NFS~\cite{haynes_network_2015}, GFS~\cite{ghemawat_google_2003}, or EFS~\cite{aws-efs}) to use a libOS, and equipping it with TLS and disk-encryption, would fail to meet all of our confidentiality, integrity, shielding, and key management requirements. Security overheads seen by these designs are also often too high to be impractical for use by any real-world application~\cite{sasy_zerotrace_2018,ahmad_obliviate_2018}.

In \BFS{}, we construct an end-to-end design from the ground up, without relying on a libOS runtime. Instead, we design a new file system core that provides protection for all data and metadata, provides comprehensive protection against host-interface attacks, and seamlessly handles encryption for persistence and transport to enable high-performance file sharing and policy management.

\section{Conclusion}
Cloud file systems have become a critical component of modern cloud infrastructure.
As threat models evolve and security requirements become stricter, new security mechanisms are needed to protect against the myriad of attacks that may be initiated by a malicious cloud provider, co-tenant, or end-client. \blue{Yet, security, functionality, and performance are often at odds with one another.} The file system must still remain flexible enough to support typical features like file sharing and policy management, and efficiently. \blue{We introduced \BFS{}, a cloud file system that satisfies all of these requirements and substantially outperforms state-of-the-art SGX-based file system designs.}
\BFS{} challenges current wisdom in cloud file system design and demonstrates that simple architectural changes can have significant practical advantages.



\bibliographystyle{IEEEtran}
\bibliography{IEEEabrv,refs}



\ifrevision\input{response}\fi

\end{document}